\documentclass[lettersize,journal]{IEEEtran}
\usepackage{bm}
\usepackage[utf8]{inputenc}
\usepackage{graphicx}
\usepackage{epstopdf}
\usepackage{multirow}
\usepackage{amsmath}
\usepackage{tabularx}
\usepackage{booktabs}
\usepackage{float}
\usepackage[linesnumbered,algoruled,boxed,lined]{algorithm2e}
\usepackage{array}
\usepackage{color}
\usepackage{amssymb}
\usepackage{cite}
\usepackage{soul}

\newtheorem{remark}{Remark}
\usepackage{adjustbox}
\usepackage{setspace}
\usepackage{makecell}
\usepackage{bbding}
\usepackage{utfsym}
\usepackage{epstopdf}
\def\BibTeX{{\rm B\kern-.05em{\sc i\kern-.025em b}\kern-.08em
		T\kern-.1667em\lower.7ex\hbox{E}\kern-.125emX}}
\normalsize

\ifCLASSINFOpdf
\else
\fi
\begin{document}
%
\title{Resource Scheduling for UAVs-aided D2D Networks: A Multi-objective Optimization Approach}

\author{\IEEEauthorblockN{Hongyang Pan, Yanheng Liu, Geng Sun,~\IEEEmembership{Member,~IEEE}, Pengfei Wang,~\IEEEmembership{Member,~IEEE}, \\Chau Yuen,~\IEEEmembership{Fellow,~IEEE}}
\vspace{-1cm}
\thanks{This study is supported in part by the National Natural Science Foundation of China (62172186, 62002133, 61872158, 62272194), in part by the Science and Technology Development Plan Project of Jilin Province (20210201072GX), and in part by China Scholarship Council. (\textit{Corresponding author: Geng Sun}).\protect}
\thanks{Hongyang Pan is with the College of Computer Science and Technology, Jilin University, Changchun 130012, China, and also with the Engineering Product Development (EPD) Pillar, Singapore University of Technology and Design, Singapore 487372 (e-mail: panhongyang18@foxmail.com).
\par Yanheng Liu and Geng Sun are with the College of Computer Science and Technology, Jilin University, Changchun 130012, China, and also with the Key Laboratory of Symbolic Computation and Knowledge Engineering of Ministry of Education, Jilin University, Changchun 130012, China (e-mail: yhliu@jlu.edu.cn, sungeng@jlu.edu.cn).
\par Pengfei Wang is with the School of Computer Science and Technology, Dalian University of Technology, Dalian 116024, China (e-mail: wangpf@dlut.edu.cn).
\par Chau Yuen is with the School of Electrical and Electronics Engineering, Nanyang Technological University, Singapore 639798 (e-mail: chau.yuen@ntu.edu.sg).
\par This manuscript has been accepted by IEEE Transactions on Wireless Communications, DOI: 10.1109/TWC.2023.3321648.
}}

\markboth{Journal of \LaTeX\ Class Files,~Vol.~18, No.~9, September~2020}%
{Submitted paper}

\maketitle

\begin{abstract}
	Unmanned aerial vehicles (UAVs)-aided device-to-device (D2D) networks have attracted great interests with the development of 5G/6G communications, while there are several challenges about resource scheduling in UAVs-aided D2D networks. In this work, we formulate a UAVs-aided D2D network resource scheduling optimization problem (NetResSOP) to comprehensively consider the number of deployed UAVs, UAV positions, UAV transmission powers, UAV flight velocities, communication channels, and UAV-device pair assignment so as to maximize the D2D network capacity, minimize the number of deployed UAVs, and minimize the average energy consumption over all UAVs simultaneously. The formulated NetResSOP is a mixed-integer programming problem (MIPP) and an NP-hard problem, which means that it is difficult to be solved in polynomial time. Moreover, there are trade-offs between the optimization objectives, and hence it is also difficult to find an optimal solution that can simultaneously make all objectives be optimal. Thus, we propose a non-dominated sorting genetic algorithm-III with a Flexible solution dimension mechanism, a Discrete part generation mechanism, and a UAV number adjustment mechanism (NSGA-III-FDU) for solving the problem comprehensively. Simulation results demonstrate the effectiveness and the stability of the proposed NSGA-III-FDU under different scales and settings of the D2D networks. 
\end{abstract}

\IEEEpeerreviewmaketitle

%
%
\section{Introduction}

\par Unmanned aerial vehicles (UAVs) have attracted much attention from both industry and academia, due to their low cost and high mobility. They have been applied extensively for disaster rescue, sensing,  and target detection \cite{DBLP:journals/corr/abs-2202-06046} \cite{DBLP:journals/tvt/GaoJXYFL22}. Recently, there are many works using UAVs in wireless networks to improve the system performance, where UAVs play various roles in the 5G/6G communications, e.g., chargers \cite{DBLP:journals/iotj/LiuP0WLL22}, aerial base stations (BSs) \cite{DBLP:journals/iotj/XuLHY21}, and aerial relays \cite{DBLP:journals/ton/ZhongGLC20}. Different from the conventional wireless communication system, UAVs can be deployed at a higher altitude and increase the probability of line-of-sight (LoS) transmissions, hence increasing the transmission gain. 

\par For air-to-ground communication scenario, UAVs are employed as aerial BSs to support wireless transmissions, so as to expand the wireless network capacity for hot spots, such as Olympics Games \cite{DBLP:conf/vtc/JavidsharifiAKS22}. Moreover, UAVs show strong advantages in emergency communications and military operations. For instance, when the military BSs are destroyed by the attack of the enemy, UAVs can be promptly utilized to offer the robust service for the ground wireless devices (WDs). Moreover, due to their high flexibility, the position deployment of the UAVs can be designed to assist the overloaded BSs in the area with dense crowds efficiently. Since the position deployment is a vital factor to influence the coverage and network performance, it is widely studied \cite{DBLP:journals/tcom/HanBBC22} \cite{DBLP:journals/icl/KaleemKMNYK22}. However, owing to the limited on-board energy, the energy consumption of the UAVs should also be considered \cite{DBLP:journals/tcom/SunLWWSL22}. 

\par Device-to-device (D2D) networks are considered as an innovative feature of the 5G/6G communications \cite{DBLP:journals/tifs/ChenLWHXZ23}. Thus, many previous works also study using UAVs as aerial relays in the D2D networks \cite{DBLP:journals/iotj/JiZNW21} \cite{DBLP:journals/tcom/AlsharoaY21}. Specifically, UAV relays can enhance the communication performance of the ground network such as achieving higher rate and lower delay by overcoming the remote distance of the ground device pairs, since the UAV-enabled aerial relays can make better use of LoS links compared to the ground relays. In addition, UAVs-aided D2D networks can be extended to several real-world applications of D2D networks. For example, in local services, user data is directly transmitted between terminals without involving the network side, such as proximity-based social media applications. Accordingly, UAV relays can greatly increase the available rate of D2D communication, and then support multimedia and other applications with high-rate requirements, so that enhancing the user experience. Moreover, for natural disasters like earthquakes, traditional communication networks based on base stations may fail to function properly due to the inflicted damage. In such cases, UAV-aided D2D networks can efficiently utilize UAV relays to achieve low-delay communication, which can further improve rescue efficiency.

\par However, introducing UAVs will cause extra flight energy consumption. Thus, how to improve the D2D network capacity in a more energy efficient way is challenging but meaningful, which motives us to propose our approach. In this work, a multiple UAVs-aided D2D network with multiple ground device pairs is investigated. Among the device pairs, those with good channel conditions and short communication distances can communicate via direct links. However, other device pairs must utilize relays since they are far apart. Thus, multiple UAVs are employed as aerial relays to forward signals. However, there are several challenges in the considered scenario. For example, the UAV positions, UAV transmission powers, communication channels, and UAV-device pair assignment can influence the network capacity largely, UAV positions and UAV flight velocities can affect flight energy consumption, and the number of deployed UAVs can also influence the D2D network capacity and flight energy consumption.

\par Different from the previous works that only consider the UAV position deployment in D2D networks, or adjust the UAV transmit powers and communication channels separately to optimize D2D network capacity, we consider to improve the D2D network capacity in a more saving way. The main contributions of this paper can be summarized as follows:

\begin{itemize}
	\item We consider a UAVs-aided D2D network, where multiple UAVs are deployed as aerial relays to serve the ground device pairs. Specifically, we formulate a UAVs-aided D2D network resource scheduling optimization problem (NetResSOP) to jointly consider the number of deployed UAVs, UAV positions, UAV transmission powers, UAV flight velocities, communication channels, and UAV-device pair assignment so as to maximize the D2D network capacity, minimize the number of deployed UAVs, and minimize the average energy consumption over all UAVs simultaneously. To the best of our knowledge, this is the first work to jointly take into account the UAV position deployment, power allocation and channel allocation of UAVs and WDs, while considering the flight energy consumption of UAVs. 

	\item Due to the continuous and discrete parts of the solution and the coupled relationship of the decision variables, the formulated NetResSOP is a mixed-integer programming problem (MIPP) and NP-hard problem. Moreover, there are trade-offs between the optimization objectives, and the dimensions of solutions are variable, which means that the conventional algorithms are difficult to solve it in polynomial time. Thus, we propose a non-dominated sorting genetic algorithm-III with a Flexible solution dimension mechanism, a Discrete part generation mechanism, and a UAV number adjustment mechanism (NSGA-III-FDU) to optimize all optimization objectives comprehensively and simultaneously. Specifically, the flexible solution dimension mechanism enables solutions with different dimensions to learn from each other, while the discrete part generation mechanism and the UAV number adjustment mechanism are used to enhance the quality of the solution.

	\item Through simulations, we verify that the proposed NSGA-III-FDU can effectively solve the formulated NetResSOP under different scales and settings. Due to the natural of the evolutionary algorithms, we provide three different alternative strategies for decision-makers, which are D2D network capacity maximum strategy (MaxNetCap), number of UAVs minimum strategy (MinUAV), and the average energy consumption minimum strategy (MinAveEnergy) according to different requirements. For MaxNetCap, we can increase the D2D network capacity by $40.50\%$ approximately, while reducing the average energy consumption over all UAVs by $5.51\%$, when using almost the same number of UAVs as comparison algorithms. For MinUAV, we can increase the D2D network capacity by $37.76\%$ approximately, while reducing the average energy consumption over all UAVs by $6.59\%$, under the condition that the number of deployed UAVs for different algorithms are almost the same. For MinAveEnergy, we can increase the D2D network capacity by $20.57\%$ approximately, while reducing the average energy consumption over all UAVs by $5.09\%$, when the gap on the number of deployed UAVs is $2.04\%$.
\end{itemize}

\par The rest of this paper is organized as follows: Section \ref{Related work} introduces the related work. The system model is given in Section \ref{System model}, while the formulation of NetResSOP is given in Section \ref{Problem statement}. Section \ref{Algorithm} proposes the algorithm, and the simulation results are shown in Section \ref{Simulation results}. Section \ref{Discussion} discusses the communication energy efficiencies with and without UAVs as well as the implementation analysis. Finally, Section \ref{Conclusion} concludes this paper.

\section{Related work}
\label{Related work}
\par In this section, 
some related works about the flight energy consumption of UAVs, position deployment of UAVs, and capacities of UAV-aided D2D networks are reviewed, and the differences between the previous works and this work are highlighted in Table \ref{Main_contributions_of_related_works}.
\begin{table*}[t]
	\setlength{\abovedisplayskip}{1pt}
	\setlength{\belowdisplayskip}{1pt}
	\setlength{\abovecaptionskip}{5pt}
	\setlength{\abovecaptionskip}{0.cm}
	\setlength{\belowcaptionskip}{-0.cm}
	\tiny
	\tabcolsep=0.05cm
	\begin{center}
		\caption{Contribution comparisons of previous works and this work}
		\begin{tabular}{|c|c|c|c|c|c|c|c|c|}
		\hline Scenario & Reference & $\begin{array}{c}\text {Multiple}\\ \text{UAVs}\\ \text {and WDs}\end{array}$ & $\begin{array}{c}\text {3D position} \\  \text{deployment} 
	\end{array}$ & $\begin{array}{l}\text { Channel } \\
				\text {allocation}\end{array}$ & $\begin{array}{c}\text {Flight energy} \\
				\text {consumption}\end{array}$ &$\begin{array}{c}\text {UAV-device}\\ \text{pair assignment}\end{array}$ & $\begin{array}{c}\text { Power } \\
				\text {allocation}\end{array}$ & $\begin{array}{c}\text { Variable} \\
				\text {UAV}\\ \text{number}\end{array}$\\
			\hline\multirow{4}{*}{$\begin{array}{c}\text {Flight energy}\\ \text {consumption}\\ \text {of UAVs}\end{array}$} & \cite{DBLP:journals/tcom/SunLWWSL22} &$\checkmark$ & $\checkmark$ & $\usym{2715}$&   $\checkmark$   & $\checkmark$& $\checkmark$  &$\usym{2715}$\\
				\cline { 2 - 9 }  &\cite{DBLP:journals/tcom/SuPCJZY22} &$\usym{2715}$ & $\checkmark$ &  $\usym{2715}$&   $\checkmark$ & $\usym{2715}$ & $\checkmark$& $\usym{2715}$\\
				\cline { 2 - 9 }  & \cite{DBLP:journals/iotj/GulEK22} &$\usym{2715}$ & $\usym{2715}$ &  $\usym{2715}$&   $\checkmark$ & $\checkmark$ & $\usym{2715}$& $\usym{2715}$\\
				\cline { 2 - 9 }   & \cite{DBLP:journals/tits/HanZZL22} &$\usym{2715}$ & $\usym{2715}$ &  $\usym{2715}$&   $\checkmark$ & $\usym{2715}$ & $\usym{2715}$& $\usym{2715}$\\
			 \hline\multirow{6}{*}{$\begin{array}{c}\text {Position}\\ \text {deployment}\\ \text {of UAVs}\end{array}$} & \cite{DBLP:journals/twc/WangDZ19} &$\checkmark$ & $\usym{2715}$ &  $\usym{2715}$&  $\usym{2715}$ & $\usym{2715}$& $\usym{2715}$& $\usym{2715}$\\
			\cline { 2 - 9} & \cite{DBLP:journals/tmc/ZhangD19} &$\checkmark$ & $\usym{2715}$ &  $\usym{2715}$&  $\usym{2715}$ & $\usym{2715}$& $\usym{2715}$& $\checkmark$\\
			\cline { 2 - 9 } & \cite{DBLP:journals/ton/ZhongGLC20} &$\checkmark$ & $\checkmark$ & $\checkmark$&  $\usym{2715}$ & $\checkmark$& $\usym{2715}$  &$\usym{2715}$\\
			\cline { 2 - 9 } & \cite{DBLP:journals/tvt/ZhongGLL20} &$\checkmark$ & $\checkmark$ & $\checkmark$&  $\usym{2715}$ & $\checkmark$& $\usym{2715}$  &$\usym{2715}$\\
			\cline { 2 - 9 } & \cite{DBLP:journals/tcom/ZhangZZZXX21} &$\checkmark$ & $\checkmark$ & $\checkmark$&  $\usym{2715}$ & $\checkmark$& $\usym{2715}$  &$\checkmark$\\
			\cline { 2 - 9 } & \cite{DBLP:journals/tcom/LiuHZZLL22} &$\checkmark$ & $\checkmark$ & $\usym{2715}$&  $\checkmark$ & $\usym{2715}$& $\usym{2715}$  &$\checkmark$\\
			\hline\multirow{4}{*}{$\begin{array}{c}\text {Capacities of}\\\text {UAV/UAVs-aided}\\ \text {D2D networks}\end{array}$} & \cite{DBLP:journals/tcom/AlsharoaY21} &$\checkmark$ & $\checkmark$ & $\usym{2715}$&  $\usym{2715}$ & $\checkmark$& $\checkmark$  &$\usym{2715}$\\
			\cline { 2 - 9 } & \cite{DBLP:journals/tgcn/ZhangOCWMC22} & $\usym{2715}$ & $\checkmark$ & $\usym{2715}$&  $\usym{2715}$ &  $\usym{2715}$& $\checkmark$  &$\usym{2715}$\\
			\cline { 2 - 9 }  & \cite{DBLP:journals/iotj/LiZZY21} &$\usym{2715}$ & $\usym{2715}$ &  $\usym{2715}$&  $\usym{2715}$ & $\checkmark$ & $\checkmark$& $\usym{2715}$\\
			\cline { 2 - 9 }  & \cite{DBLP:journals/tgcn/LiuCZWCZ21} &$\checkmark$ & $\checkmark$ &  $\usym{2715}$&  $\usym{2715}$ & $\usym{2715}$ & $\checkmark$& $\usym{2715}$\\\hline
			& This work &$\checkmark$ & $\checkmark$ & $\checkmark$&  $\checkmark$ & $\checkmark$& $\checkmark$  &$\checkmark$\\
			\hline
		\end{tabular}
		\vspace{-0.4cm}
\label{Main_contributions_of_related_works}
\end{center}
\end{table*}


\subsection{Flight energy consumption of UAVs} 
\par The flight energy consumption of UAVs is an important topic that has attracted a great of interests. For example, the authors in \cite{DBLP:journals/tcom/SunLWWSL22} considered a UAV system via collaborative beamforming to serve ground networks, so as to achieve the secure and energy-efficient communication. Specifically, they jointly reduced the effects of the eavesdroppers and minimized the propulsion energy consumption of the UAVs. However, all users could not be served simultaneously. Su \emph{et al}. \cite{DBLP:journals/tcom/SuPCJZY22} used an intelligent reflecting surface-UAV strategy to enhance the communication performance. To improve the energy efficiency of the UAV, two schemes to maximize the spectrum efficiency and energy efficiency were developed. Nonetheless, the considered scenario was based on a strong assumption that it only considered one UAV and two users. Considering a limit on the battery capacity of a UAV, energy-aware data collection was investigated in \cite{DBLP:journals/iotj/GulEK22}. Specifically, the authors proposed a UAV decision approach to obtain a better data quality among all the cluster head robots by only visiting a subset of the cluster head, while ensuring the remained energy of the UAV was not depleted. However, they still only considered one UAV in the scenario. Moreover, the authors in \cite{DBLP:journals/tits/HanZZL22} aimed to reduce the energy consumption of a UAV under UAV-assist Internet-of-Things system. Specifically, they adopted a bi-level optimization approach to optimize the UAV position deployment and trajectory, where an upper-level method was to optimize the UAV position deployment and a lower-level one was to design UAV trajectory, respectively. Nevertheless, the power allocation was ignored.

\subsection{Position deployment of UAVs}
\par The position deployment of UAVs is so vital that many previous works have considered. For example, Wang \emph{et al}. \cite{DBLP:journals/twc/WangDZ19} proposed an adaptive UAV position deployment scheme to obtain a higher average throughput, especially when the user density was not large. Then, the authors in \cite{DBLP:journals/tmc/ZhangD19} aimed to offer fast wireless coverage by using UAVs. For this purpose, several different low-complexity algorithms were proposed to solve it under different initial positions of UAVs. However, the authors in both \cite{DBLP:journals/twc/WangDZ19} and \cite{DBLP:journals/tmc/ZhangD19} only considered the one-dimension (1D) and two-dimension (2D) deployments of UAVs, which limited the flexibility of UAVs. The authors in \cite{DBLP:journals/ton/ZhongGLC20} maximized the D2D network capacity by jointly adjusting the UAV position deployment, relay assignment, and channel allocation. Then, they further extended their work to consider the trajectories of UAVs in \cite{DBLP:journals/tvt/ZhongGLL20}. Nonetheless, the flight energy consumption of UAVs was neglected in \cite{DBLP:journals/ton/ZhongGLC20} \cite{DBLP:journals/tvt/ZhongGLL20}. Zhang \emph{et al}. \cite{DBLP:journals/tcom/ZhangZZZXX21} considered a communication system assisted by multiple UAV-mounted BSs. Although they considered the three-dimension (3D) deployment of UAVs, the UAV transmission power which could influence the interference between different links was not optimized. In order to improve the quality of service of UAV-enabled wireless cellular networks, the authors in \cite{DBLP:journals/tcom/LiuHZZLL22} formulated a UAV position deployment constraint optimization problem. Nevertheless, the channel allocation which could affect the communication performance was ignored.
\subsection{Capacities of UAV-aided D2D networks}
\par Many researchers considered to improve the capacities of the UAV-aided D2D networks. For instance, the authors in \cite{DBLP:journals/tcom/AlsharoaY21} investigated a secure transmission problem in a cache-enabled UAV-relaying network, subject to an eavesdropper. Specifically, they adjusted the user association, UAV scheduling, UAV transmission power, and UAV trajectory to increase the minimum secrecy rate among users. Due to the complexity of the problem, an alternating iterative algorithm was proposed to solve the problem. However, the flight energy consumption was also ignored in this work. To extend the relaying communication duration, Zhang \emph{et al}. \cite{DBLP:journals/tgcn/ZhangOCWMC22} utilized multiple UAV cooperation and the spectral efficient UAV substitution to achieve the goal of end-to-end throughput maximization. Simulation results verified that their scheme outperforms some benchmark algorithms. Nevertheless, only one D2D pair was considered, which was not adaptive for the practical scenario. The authors in \cite{DBLP:journals/iotj/LiZZY21} studied a full-duplex UAV relaying for multiple device pairs. Specifically, a scheduling protocol exploiting time-division multiple access (TDMA) was proposed to ensure the flying flexibility of the UAV. Nonetheless, they also only considered a single UAV instead of multiple UAVs. Moreover, the authors in \cite{DBLP:journals/tgcn/LiuCZWCZ21} investigated a UAV-enabled multi-link relaying system, where multiple UAVs as aerial relays shared the same spectrum at the same time. Considering the interference between the device pairs and the UAVs, an iterative algorithm was proposed by adjusting the UAV trajectories and the transmission powers to optimize the minimum throughput of all links. However, the flight energy consumption of UAVs was not mentioned.
\subsection{Difference between the previous works and this work}
\par Different from the previous works that only considering the energy consumption during communication and UAV position deployment, we jointly consider the number of deployed UAVs, UAV positions, UAV transmission powers, UAV flight velocities, communication channels, and UAV-device pair assignment so as to maximize the D2D network capacity, minimize the number of deployed UAVs, and minimize the average energy consumption over all UAVs simultaneously. Moreover, different from the conventional methods that converting the three optimization objectives to a single optimization objective, we consider to optimize these optimization objectives by introducing the Pareto dominance, which is more practical for the considered system.

\section{System model}
\label{System model}

\par In this section, the network model, channel model, communication interference model, energy consumption model of UAV, and multi-objective optimization problem (MOP) are introduced. Moreover, Table \ref{table:notations} lists the main notations of the variables for the readability and consistency.
\begin{table*} 
	\centering
	\tiny
	\caption{List of important Notations used in this paper}
	\label{table:notations}
	\renewcommand\arraystretch{1.4}
	\begin{tabular}{|l|l|l|l|}\hline
		Variable  &Physical meanings &Variable&Physical meanings\\\hline
		
		$\mathcal{M}_S$ &SWD set of UAV-aided D2D pairs &$\mathcal{M}_S$ &DWD set of UAV-aided D2D pairs\\\hline
		
		$\mathcal{K}$ &SWD set of direct pairs &$\mathcal{C}$ &Set of Channels\\\hline
		
		$N$ &Number of UAVs &$U$ &Number of channels \\\hline
		
		$\delta(m, n)$ & Assignment relationship between SWD $m$ and UAV $n$ &$Pr_k$ & Probability of transmitting data of direct pairs  \\\hline
		
		$\epsilon(n, c_u)$ &  Assignment relationship between UAV $n$ and channel $c_u$&$\beta_0$  &Path loss at the reference distance\\\hline
		$\alpha$ &Path loss exponent &$P_m$ &Transmission power of SWD $m$\\\hline
			
		$P_n$ &Transmission power of UAV $n$ &$\gamma_{m, n}$ &SINR of UAV $n$ from SWD $m$\\\hline
		
		$I^{(S)}$ & Interference from UAVs-aided D2D pairs/UAVs &$I^{(D)}$ & Interference from direct D2D pairs\\\hline
		
		$\mathcal{H}_{c_n}$ & Set of UAVs which allocate channel $c_n$  &$\mathcal{W}_{n_*}$ & Set of SWDs which assign UAV $n_*$\\\hline
		
		$W$ & Bandwidths of all channels &$f_c$ &Center frequency of carrier\\\hline
	\end{tabular}
	\vspace{-0.4cm}
\end{table*}%
\subsection{Network model}

\par
\begin{figure*}[tbp]
	\setlength{\abovedisplayskip}{1pt}
	\setlength{\belowdisplayskip}{1pt}
	\setlength{\abovecaptionskip}{1pt}
	\setlength{\belowcaptionskip}{1pt}
	\centering{\includegraphics[width=6.5in]{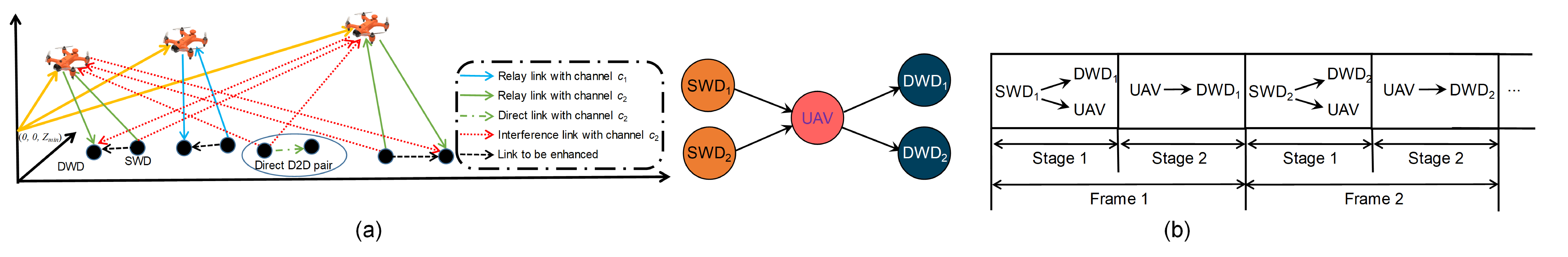}}
	\caption{System structure diagram. (a) UAVs-aided relay network model. (b) Relay working mode. }
	\label{UAVs-aided relay network}
\end{figure*}

 As shown in Fig. \ref{UAVs-aided relay network}(a), there are multiple D2D device pairs deployed on the ground with fixed locations in the considered area, and a device pair consists of a source WD (SWD) and a destination WD (DWD). Furthermore, the communication modes of the device pairs can be divided into two categories according to the difference of the communication conditions. The first category is the simple direct D2D communication, which can achieve relatively short-range LoS communication, and the pairs using this communication mode can be called as direct D2D pairs. The second category is the UAV-aided D2D communication, where one UAV serves as an aerial relay to forward signals to the DWDs. The reason for adopting this communication mode is that the communication distances between the SWDs and DWDs are relatively remote, and the links between the SWDs and DWDs will be weak. Moreover, those SWD-DWD pairs may perform more heavy works, and hence they need to be serviced for stronger information transfer. Thus, adopting UAVs as relays is to enhance the link instead of entirely neglecting the link from SWD to DWD. The pairs using this communication mode can be called as UAV-aided D2D pairs. Moreover, we consider one UAV-aided D2D pair can only be assisted by one UAV, while one UAV can be shared by multiple UAV-aided D2D pairs, and the reason that we choose this relay mode can be found in \emph{Remark} \ref{Reason for Relay mode}. The explicit working mode of a UAV is shown in Fig. \ref{UAVs-aided relay network}(b). Specifically, if a UAV is assigned by multiple SWDs, the UAV will serve those SWDs with a round-robin fashion. The form of data packet transmission is frame by frame, and the size of each frame is equal and fixed. Then, each frame can be further divided into two stages. The first stage is that one SWD transmits the data packets to the UAV and the corresponding DWD. The second stage is that the UAV forwards data packets to the corresponding DWD.
\par In this paper, we define the SWD set and the corresponding DWD set of UAV-aided D2D pairs as $\mathcal{M}_S=\{1, 2, ..., M\}$ and $\mathcal{M}_D=\{M+1, M+2, ..., 2M\}$, respectively, where $M$ is the number of UAV-aided D2D pairs, and the three-dimension (3D) coordinate of any SWD or DWD in UAV-aided D2D pairs $m\in \mathcal{M}_S \cup \mathcal{M}_D$ can be recorded as $(x_m, y_m, 0)$. The set of UAVs is defined as $\mathcal{N}=\{1, 2, ..., N\}$, where $N$ is the number of deployed UAVs, and the 3D coordinate of any UAV $n\in \mathcal{N}$ is recorded as $(x_n, y_n, z_n)$. Moreover, we assume that all UAVs start from one point, and then they will be deployed at different positions and hover for relaying. Without loss of generality, we assume that the number of deployed UAVs is smaller than the number of UAV-aided D2D pairs, i.e., $N<M$. Moreover, the SWD set of direct D2D pairs is defined as $\mathcal{K}=\{1, 2, ..., K\}$. Similarly, $K$ is the number of direct D2D pairs, and the 3D coordinate of any SWD in direct D2D pairs $k\in \mathcal{K}$ is recorded as $(x_k, y_k, 0)$. In addition, we assume the probability for any SWD $k\in \mathcal{K}$ to transmit data packets is recorded as $Pr_k$. In other words, the SWD $k\in \mathcal{K}$ is silent with a probability $1-Pr_k$. Then, the distance between WD $m^\prime\in \mathcal{M}_S \cup \mathcal{M}_D \cup \mathcal{K}$ and UAV $n\in \mathcal{N}$ is $d_{m^\prime, n}=\sqrt{(x_{m^\prime}-x_n)^2+(y_{m^\prime}-y_n)^2+{z_n}^2}$. Furthermore, we consider there are $U$ available orthogonal channels to be allocated in the considered area, and the set of them can be denoted as $\mathcal{C}=\{c_1, c_2, ..., c_U\}$, which means that all the UAVs and SWDs in direct D2D pairs can only allocate the channel from $\mathcal{C}$.  

\par Then, assume $r_m$ denotes the assigned UAV of SWD $m\in \mathcal{M}_S$, and a binary variable $\delta(m,n)$ is used to indicate the assignment relationship between SWD $m$ and UAV $n\in \mathcal{N}$, which can be expressed as follows:
\begin{equation}
\label{delta}
\delta(m,n)=\left\{
\begin{aligned}
1, \quad r_m = n \\
0, \quad r_m \neq n 
\end{aligned}
\right.
\end{equation}
\noindent Thus, we have the number of SWDs, which assign the UAV $n$, is $\mu_n=\underset{m\in \mathcal{M}_S}\sum{\delta(m,n)}$. Similarly, assume that $c_n$ is the allocated channel of UAV $n$. Then, a binary variable $\epsilon(n, c_u)$ is used to express that whether UAV $n$ allocates the channel $c_u\in \mathcal{C}$, which is as follows:
\begin{equation}
\label{epsilon}
\epsilon(n,c_u)=\left\{
\begin{aligned}
1, \quad c_n = c_u \\
0, \quad c_n\neq c_u 
\end{aligned}
\right.
\end{equation}
\begin{remark}
	\label{Reason for Relay mode} One SWD can forward signals to one DWD with several relays is extensively used in multiple-input multiple-output (MIMO) systems, and a group of UAVs can form an MIMO system by equipping multiple antennas on UAVs \cite{DBLP:journals/iotj/FengWCWGL19}. However, when the UAVs forward the signals from SWDs to DWDs with the same frequency simultaneously, the channel state information must be obtained, and the cooperation among WDs is also needed to guarantee that DWDs receive the forward signals synchronously. Such requirements are impractical to be satisfied, especially in the decentralized network. Besides, when the UAVs forward signals to a DWD following TDMA, which means that the UAVs forward the signals to the DWD at different time slots, the spectral efficiency will be reduced. Moreover, the complexity of precoding and scheduling in MIMO systems with multiple terminal WDs is high due to the mutual interference among them \cite{DBLP:journals/ton/ZhongGLC20}. Thus, we consider one UAV-aided D2D pair can only be assisted by one UAV, while one UAV can be shared by multiple UAV-aided D2D pairs. 
\end{remark} 
\subsection{Channel model}
\par In this work, we assume that the number of available channels is smaller than the number of deployed UAVs. If we denote the minimum number of deployed UAVs is $N_{min}$, we have $U<N_{min}$. Moreover, each channel is orthogonal to each other, and does not interfere with each other. Besides, the center frequency of carrier $f_c$ is the same for each channel, and each UAV and corresponding UAV-aided D2D pair can only use one frequency band, while each direct D2D pair also only use one frequency band. The abovementioned channel model means that there will be interference in the considered scenario, and the explicit details can be found in Section \ref{Communication interference model}. 
\par As mentioned above, the UAVs are used as aerial relays to forward the signals to DWDs. Thus, the communication between UAVs and WDs should use the air-to-ground channel model \cite{DBLP:journals/wcl/Al-HouraniSL14}, and the path loss between WD $m\in \mathcal{M}_S \cup \mathcal{M}_D \cup \mathcal{K}$ and UAV $n \in \mathcal{N}$ can be expressed as follows:
\begin{equation}
\label{A2G}
\begin{split}
PL_{m,n}=&\frac{\eta_{LoS}-\eta_{NLoS}}{1+a\exp[-b(\frac{180}{\pi}\theta_{m, n}-a)]}\\&+20\log \left(\frac{4\pi f_c d_{m, n}}{c}\right)+\eta_{NLoS}
\end{split}
\end{equation}
\noindent where $a$, $b$, $\eta_{LoS}$ and $\eta_{NLoS}$ are model parameters which can be decided according to the communication environment. $\theta_{m, n}=\sin^{-1}(\frac{z_n}{d_{m,n}})$ is the elevation angle in degree from WD $m$ to UAV $n$. Then, the channel gain between WD $m$ and UAV $n$ can be expressed as follows:
\begin{equation}
\label{gain-A2G}
\begin{split}
h_{m,n}=10^{-PL_{m,n}/10}
\end{split}
\end{equation}
\par However, the communication between two WDs on the ground is not applicable to the abovementioned channel model. Thus, we use the typical LoS model for such a situation, and the gain between a WD $k$ and another WD $k^\prime$ can be expressed as follows:
\begin{equation}
\label{gain-G2G}
\begin{split}
h^\prime_{k, k^\prime}=\beta_0 d_{k, k^\prime}^{-\alpha}
\end{split}
\end{equation}
\noindent where $\beta_0$ is the path loss at the reference distance, and $\alpha$ is the path loss exponent.
\subsection{Communication interference model}
\label{Communication interference model}
\par In this paper, multiple UAVs are deployed as aerial relays to forward signals to the DWDs, subject to direct D2D pairs. Let $P_m$ represent the transmission power of SWD $m$, and then the instantaneous signal to interference plus noise ratio (SINR) of UAV $n$ from SWD $m$ can be expressed as follows:
\begin{equation}
\label{SINRmn}
\begin{split}
\gamma_{m,n}=\frac{P_mh_{m, n}}{\sigma^2+I_{m,n}}
\end{split}
\end{equation}
\noindent where $\sigma^2$ is the white Gaussian noise power, and $I_{m,n}$ is the interference of UAV $n$ when it receives the signal from SWD $m$, respectively. Specifically, $I_{m,n}$ can be divided into two parts. First, it is the interference from SWDs of UAV-aided D2D pairs that allocate the same channel as UAV $n$, which can be recorded as $I_{m,n}^{(S)}$. Second, it is the interference from SWDs of direct D2D pairs that allocate the same channel as UAV $n$, which can be recorded as $I_{m,n}^{(D)}$. Then, $I_{m,n}=I_{m,n}^{(S)}+I_{m,n}^{(D)}$. As can be seen, both of $I_{m,n}^{(S)}$ and $I_{m,n}^{(D)}$ are the instantaneous values, thus $I_{m,n}$ is also an instantaneous value. However, the value of $I_{m,n}^{(S)}$ is time-varying, and $I_{m,n}^{(D)}$ does not always have the value since SWDs of direct D2D pairs send data packets probabilistically. Thus, the exact form of $I_{m,n}$ is hard to express so that the expected form of $\gamma_{m,n}$ is used in this paper, which can be expressed approximatively as follows:
\begin{equation}
\label{SINRmn_approximatively}
\begin{split}
\bar{\gamma}_{m,n}=\frac{P_mh_{m, n}}{\sigma^2+\textbf{E}[I_{m,n}]}
\end{split}
\end{equation}
\noindent where $\textbf{E}[\cdot]$ is the expectation of $[\cdot]$.  $\textbf{E}[I_{m, n}]$ is the expected interference from SWD $m$ received by UAV $n$ which can be further expressed as follows:  
\begin{equation}
\label{EImn}
\begin{split}
\textbf{E}[I_{m, n}]=\textbf{E}[I_{m, n}^{(S)}]+\textbf{E}[I_{m, n}^{(D)}]
\end{split}
\end{equation}
\par Let $\mathcal{H}_{c_n}$ represent the set of UAVs which allocate channel $c_n$. Then, the SWDs which assign UAV $n_*\in \{\mathcal{H}_{c_n} \backslash n\}$ may interfere UAV $n$ if SWD $m$ transmits signals to UAV $n$. Assume $\mathcal{W}_{n_*}$ is the set of SWDs, which assign UAV $n_*$, and $|\mathcal{W}_{n_*}|$ is the number of SWDs in $\mathcal{W}_{n_*}$. Thus, the expected interference from SWDs of UAV-aided D2D pairs can be expressed as follows: 
\begin{equation}
\label{EImnS}
\begin{split}
\textbf{E}[I_{m, n}^{(S)}]=\underset{n_*\in \{\mathcal{H}_{c_n} \backslash n\}}\sum\underset{w_i \in \mathcal{W}_{n_*}}\sum \frac{P_{w_i}h_{w_i, n}}{|\mathcal{W}_{n_*}|}
\end{split}
\end{equation}
\noindent Since SWD $k$ transmits data packets with a probability $Pr_k$, the expected interference from SWDs of direct D2D pairs can be expressed as follows:
\begin{equation}
\label{EImnD}
\begin{split}
\textbf{E}[I_{m, n}^{(D)}]=\underset{k \in \mathcal{K}_{c_n}}\sum Pr_k P_kh_{k, n}
\end{split}
\end{equation}
\noindent where $P_k$ is the transmission power, and $\mathcal{K}_{c_n}$ represents the set including SWDs of direct D2D pairs that allocate channel $c_n$. Then, according to Eqs. (\ref{EImnS}) and (\ref{EImnD}), Eq. (\ref{EImn}) can be rewritten as follows:
\begin{equation}
\label{EImnrewritten}
\begin{split}
\textbf{E}[I_{m, n}]=\underset{n_*\in \{\mathcal{H}_{c_n} \backslash n\}}\sum\underset{w_i \in \mathcal{W}_{n_*}}\sum \frac{P_{w_i}h_{w_i, n}}{|\mathcal{W}_{n_*}|}+\underset{k \in \mathcal{K}_{c_n}}\sum Pr_k P_kh_{k, n}
\end{split}
\end{equation}

\par Similarly, assume that $P_n$ is the transmission power of UAV $n$, and then the expected SINR of $(m+M)$-th DWD from UAV $n$ can be indicated as follows:
\begin{equation}
\label{SINRnm+M_approximatively}
\begin{split}
\bar{\gamma}_{n, m+M}=\frac{P_nh_{n, m+M}}{\sigma^2+\textbf{E}[I_{n, m+M}]}
\end{split}
\end{equation}
\noindent where $\textbf{E}[I_{n, m+M}]$ is expected interference of DWD $m+M$ from UAV $n$, which can also be divided two parts: first, it is the interference from other UAVs using the channel $c_n$. Second, it is the interference from SWDs of direct D2D pairs that allocate the channel $c_n$. Since the DWDs of UAV-aided D2D pairs and direct D2D pairs are all on the ground, the latter interference should adopt typical LoS channel. Thus, the explicit expression of $\textbf{E}[I_{n, m+M}]$ is as follows:
\begin{equation}
\label{EInm+M}
\begin{split}
\textbf{E}[I_{n, m+M}]=\underset{n_*\in \{\mathcal{H}_{c_n} \backslash n\}}\sum P_nh_{n*,m+M}+\underset{k \in \mathcal{K}_{c_n}}\sum Pr_k P_kh^\prime_{k, m+M}
\end{split}
\end{equation}
\par Since the SWDs and DWDs of direct D2D pairs are all on the ground, the expected SINR of the direct transmission from SWD $m$ to DWD $m+M$ is as follows:
\begin{equation}
\label{SINRmm+M_approximatively}
\begin{split}
\bar{\gamma}_{m, m+M}=\frac{P_nh^\prime_{m, m+M}}{\sigma^2+\textbf{E}[I_{m, m+M}]}
\end{split}
\end{equation}
\noindent where $\textbf{E}[I_{m, m+M}]$ is the expected interference received by DWD $m+M$ from SWD $m$, which can be indicated as follows:
\begin{equation}
\label{EImm+M}
\begin{split}
\textbf{E}[I_{m, m+M}]=&\underset{n_*\in \{\mathcal{H}_{c_n} \backslash n\}}\sum\underset{w_i \in \mathcal{W}_{n_*}}\sum \frac{P_{w_i}h_{w_i, m+M}}{|\mathcal{W}_{n_*}|}+\\&\underset{k \in \mathcal{K}_{c_n}}\sum Pr_k P_kh^\prime_{k, m+M}
\end{split}
\end{equation}

\subsection{Energy consumption model of UAV}
\label{Energy consumption model of UAV}

\par For a rotary-wing UAV, since motion energy consumption is about dozens of times that of communication energy consumption \cite{DBLP:journals/twc/ZengXZ18}, we only consider the motion energy consumption in this paper. Thus, the energy consumption model in two-dimension horizontal space of a rotary-wing UAV is adopted referred to \cite{DBLP:journals/twc/ZengXZ19}, which is as follows:
\begin{equation}
\label{UAV-2D-Power}
\begin{split}
P(V)=&P_{B}\left(1+\frac{3{V}^2}{U_{tip}^{2}}\right)+P_{I}\left(\sqrt{1+\frac{{V}^4}{4v_{0}^{4}}}-\frac{{V}^2}{2v_{0}^{4}}\right)^{\frac{1}{2}}\\&+\frac{1}{2}d_{0}\rho sAV^{3}
\end{split}
\end{equation}

\noindent where $P_B$ and $P_I$ are the blade profile power and the induced power in hovering status, respectively, which are both constants. $U_{tip}$ and $v_{0}$ represent the tip speed of the rotor blade and the average rotor induction speed hovering, respectively. $d_{0}$, $\rho$, $s$ and $A$ are the airframe drag ratio, the air density, the rotor solidity and the area of the rotor disk, respectively. Moreover, we ignore the effect from the acceleration or deceleration of UAV since it costs a tiny time in the total \cite{DBLP:journals/twc/ZengXZ19}.

\par Moreover, this model is extended in 3D space by considering the UAV climbing and descending over tie, and the energy consumption model can be approximately expressed as follows \cite{DBLP:journals/pieee/ZengWZ19}: 
\begin{equation}
\label{UAV-3D-Power}
\begin{split}
E(V)\approx &\int_{0}^{T} P(V(t))dt+\frac{M_{UAV}({V(T)}^2-{V(0)}^2)}{2}\\&+M_{UAV}g(Z(T)-Z(0))
\end{split}
\end{equation}
\noindent where $V(t)$ represents the instantaneous UAV speed at time $t$, and $T$ is the total flight time. $M_{UAV}$ and $g$ represent the mass and the gravitational factor, respectively. $Z(T)-Z(0)$ represents the altitude change of a UAV from the initial position to the deployed position.
%

\subsection{Multi-objective optimization problem}
\label{Multi-objective optimization problem}
\begin{figure}[t]
	\centering{\includegraphics[width=3in]{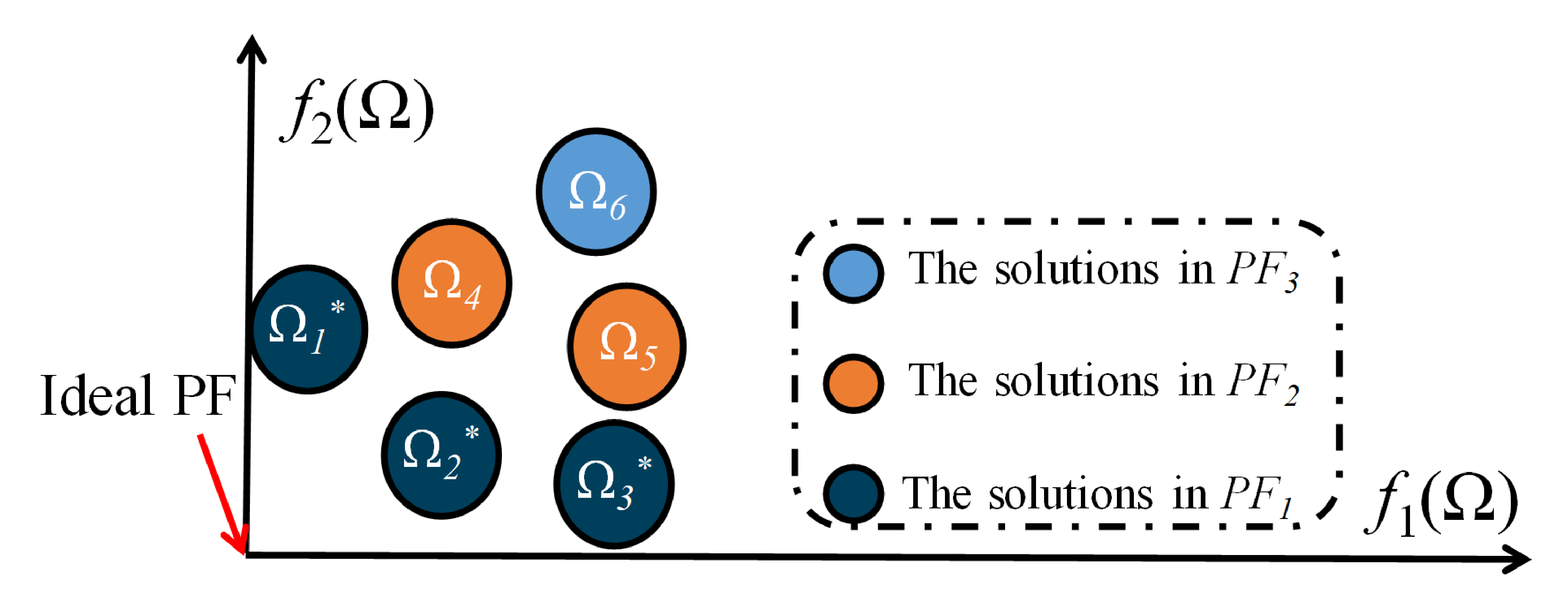}}
	\caption{Example diagram of an MOP with two optimization objectives.}
	\label{MOP}
\end{figure}

In an MOP, it is difficult to directly compare the value of single optimization objective, since the following situations may occur when directly comparing two solutions, i.e., $f_1(\Omega_1)<f_1(\Omega_2)$ but $f_2(\Omega_1)>f_2(\Omega_2)$. More intuitively, Fig. \ref{MOP} shows an example of MOP with two optimization objectives. Specifically, $\Omega_4$ and $\Omega_5$ are two solutions which satisfy the abovementioned relationship. Moreover, we define $\Omega_6$ is dominated by $\Omega_4$, recorded as $\Omega_4\prec \Omega_6$, if and only if $\Omega_4$ is better than $\Omega_6$ in at least one objective, and the mathematical expression is as follows:
\begin{subequations}
	\label{eqMOP}
	\begin{align}
		f_q(\Omega_4)\leq f_q(\Omega_6), \forall q\in {1, 2, 3,...,Q}\\
		f_q(\Omega_4)< f_q(\Omega_6), \exists q\in {1, 2, 3,...,Q}		
	\end{align}
\end{subequations}

\noindent where $Q$ is the number of objective functions. $\Omega^*$ is the non-dominated solution, such as $\Omega^*_1$, $\Omega^*_2$ and $\Omega^*_3$ in Fig. \ref{MOP}, if there is no $\Omega^*$ satisfying $\Omega\prec \Omega^*$. The non-dominated solution set contains all $\Omega^*$, usually has multiple solutions with non-dominance instead of a single solution, and the corresponding objective space of non-dominated solution set is the Pareto front (PF). Moreover, we define that the PF with optimal values for all objectives is the ideal PF, although the ideal PF is difficult to obtain due to the trad-offs between the optimization objectives.

\section{NetResSOP formulation based on Multi-objective optimization}
\label{Problem statement}

\par In this work, we consider a square horizontal area, and the minimum and maximum horizontal range of the deployed UAV is $L_{min}$ and $L_{max}$. Assume that the initial position of all UAVs is $O=(0, 0, Z_{min})$, and they can be deployed between $[Z_{min}, Z_{max}]$. Since the UAVs have certain distances from WDs, it may cause the low SINR so as to reduce the capacity of D2D network. However, using a single UAV to fly close to the WDs for communication is impractical, since the flight energy consumption of the UAV will be increased. Thus, it is better to deploy multiple UAVs to aid the D2D network, so that achieving a greater D2D network capacity. However, there are several challenges for the considered deployment. \textbf{First}, there is the mutual interference among the SWDs that assign the same UAV, and the mutual interference also exists among the UAVs that allocate the same channel. \textbf{Second}, although deploying the UAVs, which are closer to WDs, can improve the local SINR, the global D2D network capacity may be influenced, since some WDs cannot communicate with a UAV effectively. \textbf{Finally}, the position deployment of the UAVs will generate extra flight energy consumption, and the UAV flight velocities from the initial position to deployed position can also influence the flight energy consumption, which influence the lifetime of the UAVs.

\par In order to solve the abovementioned challenges, we formulate an NetResSOP to maximize the D2D network capacity, minimize the number of deployed UAVs, and minimize the average energy consumption over all UAVs simultaneously. Furthermore, we comprehensively and jointly consider the number of deployed UAVs, UAV positions, UAV transmission powers, UAV flight velocities, communication channels, and UAV-device pair assignment, since all of them can influence the optimization results. Specifically, the abovementioned three optimization objectives are as follows.

\noindent \emph{\textbf{Optimization objective 1: maximize the D2D network capacity.}} Assume that the amplify-and-forward protocol is used for relaying \cite{DBLP:journals/tit/LanemanTW04}, and the bandwidth of the channel is $W$. Then, the expected transmission rate of the link from $m$ to $m+M$ via UAV $n$ is expressed as follows:
\begin{equation}
\label{R_ave}
\begin{split}
 &\bar{R}_{m,n} =\frac{W\cdot \delta(m,n)}{2\mu_n} \\&\log_2[1+\bar{\gamma}_{m,m+M}+\bar{\gamma}_{m,n}\bar{\gamma}_{n,m+M}/(1+\bar{\gamma}_{m,n}+\bar{\gamma}_{n,m+M})]
\end{split}
\end{equation}
\noindent The expected capacity of SWD $m$ and UAV $n$ can be expressed as follows:
\begin{subequations}
\label{R_mR_n}
\begin{align}
\bar{R}_{m}= \underset{n\in \mathcal{N}}\sum \bar{R}_{m, n}\\
\bar{R}_{n}= \underset{m\in \mathcal{M}_S}\sum \bar{R}_{m, n}
\end{align}
\end{subequations}
\par Then, the first optimization objective, i.e., the expected D2D network capacity can be expressed as follows \cite{DBLP:journals/ton/ZhongGLC20}:
\begin{equation}
\label{f_1}
\begin{split}
f_1(\mathbb{X}^{1\times N}, \mathbb{Y}^{1\times N}, \mathbb{Z}^{1\times N}, \mathbb{P}^{1\times N}, \mathbb{A}^{M\times N}, \mathbb{B}^{(N+K)\times U}, N) \\= \underset{m\in \mathcal{M}_S}\sum \underset{n\in \mathcal{N}}\sum \bar{R}_{m,n}
\end{split}
\end{equation}

\noindent where $[\mathbb{X}^{1\times N}, \mathbb{Y}^{1\times N}, \mathbb{Z}^{1\times N}, \mathbb{P}^{1\times N}, \mathbb{A}^{M\times N}, \mathbb{B}^{(N+K)\times U}, N]$ is the solution of $f_1$. Specifically, $[\mathbb{X}^{1\times N}, \mathbb{Y}^{1\times N}, \mathbb{Z}^{1\times N}]$ are the deployed positions of the UAVs, and $\mathbb{P}^{1\times N}$ are the transmission powers of the UAVs. $\mathbb{A}^{M\times N}$ is the set of $\delta$ which represents the UAV-device pair assignment relationship, $\mathbb{B}^{(N+K)\times U}$ is the set of $\epsilon$ which represents the channel allocation relationship of UAVs and the direct D2D pairs, and $N$ is the number of deployed UAVs.

\noindent \emph{\textbf{Optimization objective 2: minimize the number of deployed UAVs.}} To obtain the maximum D2D network capacity, we can employ more UAVs certainly. However, the number of available UAVs is often limited in the practical application. Moreover, more UAVs will cause the increasing of the total energy consumption, which leads to unnecessary resource waste. In addition, too few UAVs can increase the mutual interference if they allocate the same channel. Thus, we aim to optimize the number of deployed UAVs, while obtaining the maximum D2D network capacity. The second optimization objective is to minimize the number of deployed UAVs, which can be expressed as follows:
\begin{equation}
\label{f_2}
\begin{split}
f_2(\mathbb{X}^{1\times N}, \mathbb{Y}^{1\times N}, \mathbb{Z}^{1\times N}, N)= N
\end{split}
\end{equation}

\noindent \emph{\textbf{Optimization objective 3: minimize the average energy consumption over all UAVs.}} In order to optimize the D2D network capacity, each UAV needs to move to the reasonable position for achieving relaying. The moving process can generate extra flight energy consumption undoubtedly, and it should be as less as possible since each UAV has a limited on-board energy. As mentioned above, the communication energy consumption is ignored since it is much smaller than flight energy consumption, and hence it is more practical to consider the flight energy consumption for the real-world scenario. Assume that UAV $n$ flies with a constant speed $V_n$, the flight time of UAV is $T_n=D_n/V_n$, where $D_n$ is the distance between the deployed position of UAV $n$ and the initial position $O=(0,0,Z_{min})$. Note that the velocities of the different UAVs are allowed to be different. The flight energy consumption of UAV $n$ can be calculated according to Eq. (\ref{UAV-3D-Power}), which can be recorded as $E_n$. Then, the third optimization objective is to minimize the average energy consumption over all UAVs, which can be expressed as follows:
\begin{equation}
\label{f_3}
\begin{split}
f_3(\mathbb{X}^{1\times N}, \mathbb{Y}^{1\times N}, \mathbb{Z}^{1\times N}, \mathbb{V}^{1\times N}, N)=  {\underset{n\in \mathcal{N}}\sum {E}_{n}/N}
\end{split}
\end{equation}

\begin{figure}[tbp]
	\centering{\includegraphics[width=2.5in]{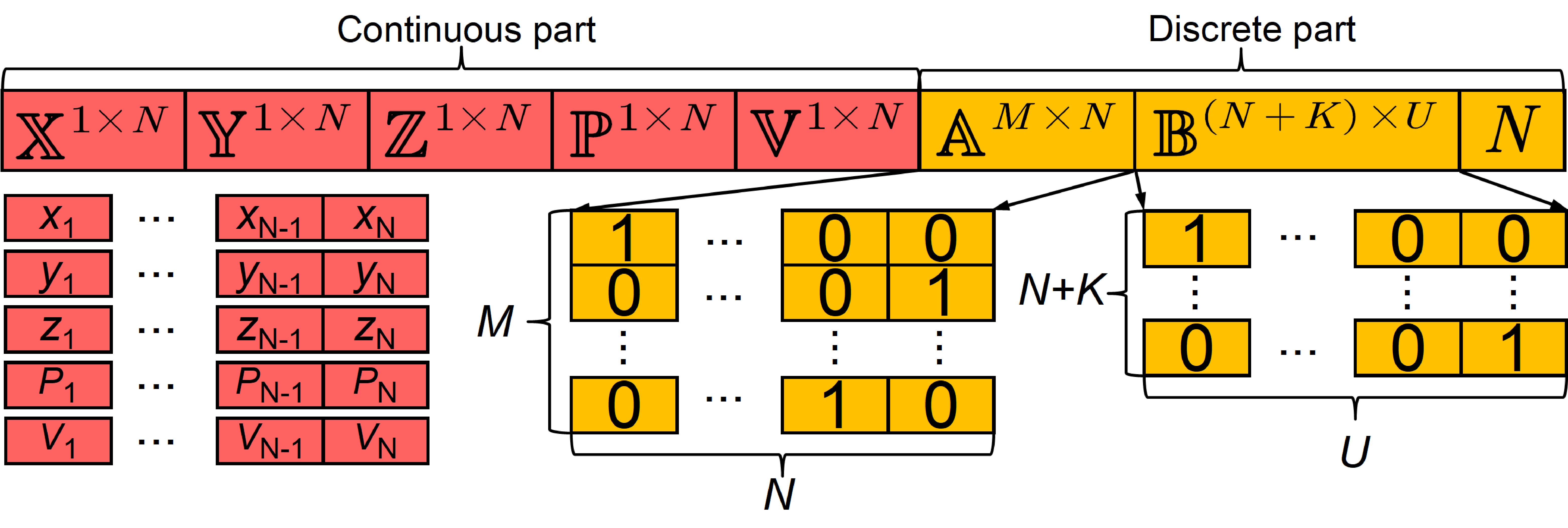}}
	\caption{Solution structure of the formulated NetResSOP.}
	\label{Solution structure}
\end{figure}

\noindent Moreover, to restrict the flight time interval between the first and last arrival of the UAVs at the corresponding deployment positions so that ensuring the interval cannot be too large, we set a time threshold $T_{th}$, which aims to enable simultaneous deployment and recall of UAVs as much as possible. Once one of UAVs is going to run out of energy, all UAVs will be recalled simultaneously. The time threshold constraint can be expressed as follows:
\begin{equation}
	\label{Time_Th}
	\begin{split}
		\widetilde{\text{Max}}\{T_n\}-\widetilde{\text{Min}}\{T_n\}\leq T_{th}
	\end{split}
\end{equation}

\noindent where $\widetilde{\text{Max}}\{\cdot\}$ and $\widetilde{\text{Min}}\{\cdot\}$ represent taking the maximum and minimum from the set, respectively. Note that $f_1$ is to obtain the maximum while $f_2$ and $f_3$ are to obtain the minimum. Thus, to unify the direction of optimization, we take the negative value of $f_1$ to obtain the minimum of $-f_1$, $f_2$ and $f_3$, in which the minimum of $-f_1$ corresponds the maximum of $f_1$. Then, the NetResSOP can be formulated as follows:
\begin{subequations}
	\label{NetResSOP}
	\begin{align}
	{\underset{\mathbf{X}}{\text{min}}}  \quad  & F=\{-f_{1}, f_{2}, f_{3} \}\\
	\text{s.t.} \quad &C1: L_{min} \leqslant  x_{n} \leqslant  L_{max}, \forall n \in \mathcal{N} \\
	&C2: L_{min} \leqslant  y_{n} \leqslant  L_{max}, \forall n \in \mathcal{N}\\
	&C3: Z_{min} \leqslant  z_{n} \leqslant  Z_{max}, \forall n \in \mathcal{N}\\
	&C4: P_{min} \leqslant  P_{n} \leqslant  P_{max}, \forall n \in \mathcal{N}\\  
	&C5: V_{min} \leqslant  V_{n} \leqslant  V_{max}, \forall n \in \mathcal{N}\\    
	&C6: \underset{n\in \mathcal{N}}\sum \delta(m,n)=1, \forall m\in\mathcal{M}_S\\
	&C7:\underset{c_u\in \mathcal{C}}\sum \epsilon(n,c_u)=1, \forall n\in\mathcal{N}\\
	&C8:\underset{c_u\in \mathcal{C}}\sum \epsilon(k,c_u)=1, \forall k\in\mathcal{K}\\
	&C9: N_{min} \leq N \leq N_{max}\\
	&C10:\widetilde{\text{Max}}\{T_n\}-\widetilde{\text{Min}}\{T_n\}\leq T_{th}
	\end{align}
\end{subequations}
\noindent where $\mathbf{X}=[\mathbb{X}^{1\times N}, \mathbb{Y}^{1\times N}, \mathbb{Z}^{1\times N}, \mathbb{P}^{1\times N}, \mathbb{V}^{1\times N}, \mathbb{A}^{M\times N},\\ \mathbb{B}^{(N+K)\times U}, N]$ is the whole solution of the formulated NetResSOP, whose explicit form is shown in Fig. \ref{Solution structure}. $[\mathbb{X}^{1\times N}, \mathbb{Y}^{1\times N}, \mathbb{Z}^{1\times N}, \mathbb{P}^{1\times N}, \mathbb{V}^{1\times N}]$ is continuous part of the solution, while $[\mathbb{A}^{M\times N}, \mathbb{B}^{(N+K)\times U}, N]$ is the discrete part of the solution.  $C1 - C5$ limit the bounds of the continuous part of the solution, while $C6 - C9$ limit the domains of the discrete part of the solution. Specifically, $C6$ is to ensure that each UAV-aided D2D pair can be assigned to only one UAV as its relay. Similarly, $C7$ and $C8$ are to limit that one UAV and one direct D2D pair can be allocated to only one channel, respectively. $C9$ is to limit the number of deployed UAVs. $C10$ is to limit that the difference of maximum and minimum deployed times cannot exceed the preset threshold.

\begin{remark} As can be seen, there are both continuous and discrete parts in the solution. Thus, the formulated NetResSOP is an MIPP \cite{DBLP:journals/candie/MengZRZL20}, and it is NP-hard \cite{nair2020solving}. In addition, note that $C6 - C8$ define all-different constraints. For ease of analysis, we only analyze $C6$. The explicit form of $\mathbb{A}^{M\times N}$ is shown in Fig. \ref{Solution structure}, which is the set of $\delta(m,n)$. For example, the second row of $\mathbb{A}^{M\times N}$ in Fig. \ref{Solution structure} expresses that $N$-th UAV is assigned by the second UAV-aided D2D pair. Moreover, the decision variables of a solution are coupled, which means that one decision variable change may influence the optimization results of other decision variables, which will further increase the difficulty of the problem. Additionally, there are trade-offs between the three optimization objectives. On the one hand, increasing the number of deployed UAVs will improve the communication performance. On the other hand, the flight energy consumption may be sacrificed for better communication performance. In other words, it is difficult to find an optimal solution that can simultaneously make all objectives be optimal. To overcome the trade-offs and optimize the three optimization objectives simultaneously, most of the decision makers identify a linear weight based on a pre-given trade-off to optimize all optimization objectives, whereas the weight is difficult to determine, since the positions of WDs may be different in various scenarios. Therefore, it is unrealistic to require requestors to determine a satisfying trade-off between multiple objectives a priori \cite{DBLP:journals/tmc/WangYHGX18}. Accordingly, we introduce the Pareto dominance to compare different solutions. 
\end{remark}
%
%
\section{Proposed NSGA-III-FDU}

\label{Algorithm}
\par The formulated NetResSOP is an MIPP and an NP-hard problem, and hence it is hard to solve in polynomial time. Moreover, it is difficult to use a convex or non-convex optimization methods, such as successive convex approximation, to solve it. Besides, if we utilize reinforcement learning to solve the formulated NetResSOP, the optimization objectives will be converted as a single reward function \cite{9777886}, which will reduce the solution space and fall into local dilemma. Evolutionary algorithms, such as non-dominated sorting genetic algorithm-III (NSGA-III), can obtain a feasible solution in polynomial time \cite{zhao2023self} \cite{DBLP:journals/iotj/LiuWSL22}. 

\par The abovementioned reasons motive us to use evolutionary algorithms with several improved specific designs to solve the formulated problem. Moreover, decision-makers will obtain a set of feasible solutions instead of a single solution using multi-objective evolutionary algorithms, and they can select the appropriate solution according to the realistic requirements. Among many evolutionary algorithms, NSGA-III performs superiority for dealing with the constraint programming problem with multiple optimization objectives \cite{DBLP:journals/tec/DebJ14} \cite{DBLP:journals/tec/JainD14}. However, the conventional NSGA-III is generally used for the problem with only continuous solution, and hence it cannot solve the formulated NetResSOP as the solution of NetResSOP contains discrete part. Moreover, the dimensions of the solutions in the conventional NSGA-III are fixed and equal. However, the number of deployed UAVs is included in the solution, which means that the dimensions of two solutions with different UAV numbers are different. Such two solutions will cause that the conventional NSGA-III cannot achieve crossover and mutation. Thus, the abovementioned factors motive us to propose an improved NSGA-III-FDU based on the conventional NSGA-III to solve the formulated NetResSOP better, and the details are as follows.
\subsection{Conventional NSGA-III}

\par The basic framework of the conventional NSGA-III is similar to the non-dominated sorting genetic algorithm-II (NSGA-II) excluding the selection operator. However, the population diversity of the NSGA-III is maintained by supplying and updating a great number of well-spread reference points instead of crowding distance. Specifically, assume $\mathcal{P}_{it}$ represents the $it$-th iteration parent population and the population size is set as $Pop$. First, the conventional NSGA-III utilizes the simulated binary crossover (SBX) \cite{DBLP:journals/tec/DebAPM02} and polynomial mutation (PM) \cite{DBLP:journals/tec/DebAPM02} to generate the offspring population $\mathcal{Q}_{it}$. Second, combine the parent population $\mathcal{P}_{it}$ and the offspring population $\mathcal{Q}_{it}$ to obtain the population $\mathcal{R}_{it}=\mathcal{P}_{it}\cup\mathcal{Q}_{it}$ (of size $2Pop$). Finally, use the non-dominated sorting mechanism and reference point mechanism jointly to generate the $(it+1)$-th parent population  $\mathcal{P}_{it+1}$, so as to maintain the population size. Specifically, the non-dominated sorting mechanism in the conventional NSGA-III is the same as NSGA-II, i.e., $\mathcal{R}_{it}$ can be ranked as several levels, which can be recorded as $\mathcal{F}_1$, $\mathcal{F}_2$, ..., etc. Add $\mathcal{F}_1$, $\mathcal{F}_2$, ..., $\mathcal{F}_l$ to construct a new population $\mathcal{S}_{it}$ until the population size of $\mathcal{S}_{it}$ is equal to $Pop$ or for the first time larger than $Pop$. There are two situations needed to be considered. On the one hand, if $|\mathcal{S}_{it}|=Pop$, go directly to the next iteration with $\mathcal{P}_{it+1}=\mathcal{S}_{it}$. On the other hand, if $|\mathcal{S}_{it}|>Pop$, $\mathcal{P}_{it+1}=\cup_1^{l-1}{\mathcal{F}}_{l-1}$. Besides, there are still $(Pop-\sum_1^{l-1}|{\mathcal{F}}_{l-1}|)$ elements from $\mathcal{F}_l$ that need to be added in  $\mathcal{P}_{it+1}$ according to the reference point mechanism, which can be found in \cite{DBLP:journals/tec/DebJ14}.

\subsection{Proposed NSGA-III-FDU}

\par In this section, an improved NSGA-III-FDU is proposed to solve the formulated NetResSOP. As mentioned above, NetResSOP is a difficult problem and the conventional NSGA-III has a few advantages for solving it. However, the conventional NSGA-III is hard to solve the formulated NetResSOP directly, and the reasons are as follows: \textbf{first}, since the dimension of the solution is not fixed, a flexible solution dimension mechanism is essential to make the algorithm execute SBX and PM. \textbf{Second}, due to the simultaneous existence of continuous and discrete parts of the solution, a discrete part generation mechanism is necessary to deal with the discrete part. \textbf{Finally}, as shown in Fig. \ref{Solution structure}, a solution consists of several parts with different physical meanings, which means the bounds of them are different. It is better to utilize different methods to update them respectively. Thus, we introduce a UAV number adjustment mechanism to update the UAV number of the solution, so as to improve the performance of the algorithm. Assume that $\mathcal{P}_{it}$ is the population of the $it$-th iteration, and $Pop$ is the population size. Then, the pseudo-code of the NSGA-III-FDU is shown in Algorithm \ref{NSGA-III-FDU}, and the improved factors are detailed as follows: 

\begin{algorithm}[t]
	\caption{NSGA-III-FDU}
	\label{NSGA-III-FDU}
	{\textbf{Input:} $Pop$, $G_{max}$, $\mathcal{P}_{0}$, etc.\\
	 \textbf{Output:} the first Pareto dominated set $\mathcal{F}^{G_{max}}_{1}$.\\	
	 	$\mathcal{P}_{0}\Leftarrow \varnothing$;\\
	 	\For{$pop$= $1$ to $Pop$}{
	 	Initialize the continuous part and discrete part randomly;\\
	 	Generate the discrete part using \emph{\textbf{Algorithm}} \ref{Random_search_operator} and then update $\mathcal{P}_{0}$;\\
	 	Execute the flexible solution dimension mechanism;\\
 		}
 		\For{$it$= $1$ to $G_{max}$}{
 			\For{$pop$= $1$ to $Pop$}{
 				Generate the offspring $\mathcal{Q}_{it}$ using SBX \cite{DBLP:journals/tec/DebAPM02} and PM \cite{DBLP:journals/tec/DebAPM02}, $\mathcal{Q}^\prime_{it}\Leftarrow \mathcal{Q}_{it}$;\\
 				Update the discrete part of $\mathcal{Q}_{it}$ using \emph{\textbf{Algorithm}} \ref{Probabilistic_learning_operator};\\ 
 				Update the UAV number of $\mathcal{Q}^\prime_{it}$ according to \emph{\textbf{Algorithm}} \ref{UAVrelayadjustment};\\
 				Update the UAV-device pair assignment and channel allocation of $\mathcal{Q}^\prime_{it}$ using \emph{\textbf{Algorithm}} \ref{Random_search_operator};\\
 				Calculate the objective functions according to Eqs. (\ref{delta})-(\ref{NetResSOP});\\
 				$\mathcal{P}_{it+1}\Leftarrow \mathcal{P}_{it} \cup\mathcal{Q}_{it}\cup\mathcal{Q}^\prime_{it}$;\\
 				Remove the redundant solutions from $\mathcal{P}_{it+1}$ to maintain the population size according to the conventional NSGA-III;\\		
 			}
 		}
	}	
\end{algorithm}
\subsubsection{Flexible solution dimension mechanism} 
\begin{figure}[tbp]
	\centering{\includegraphics[width=3.5in]{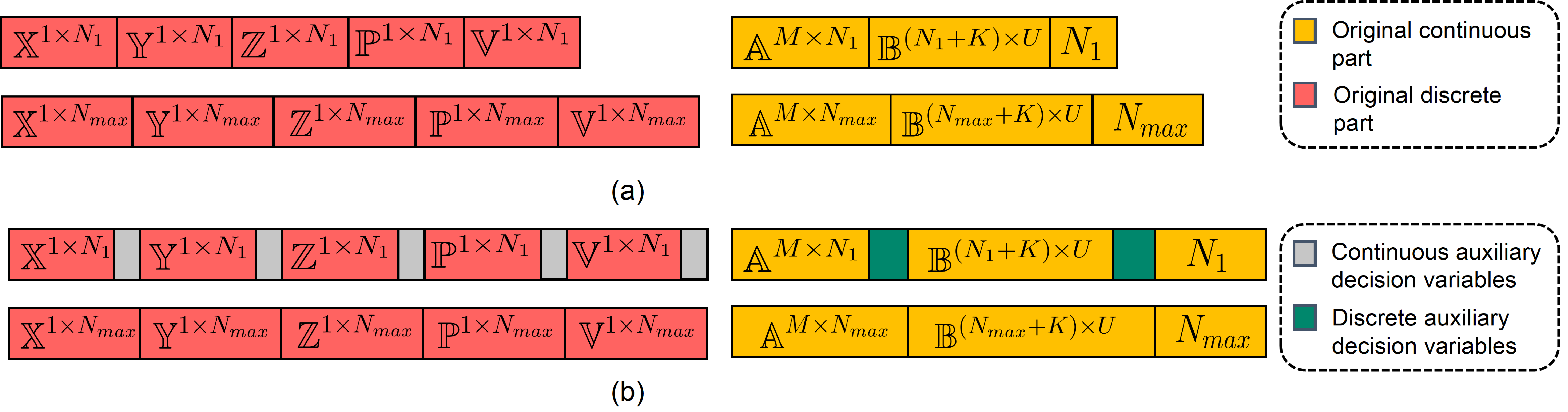}}
	\caption{Basic idea of flexible solution dimension mechanism. (a) Original population. (b) The whole population after flexible solution dimension mechanism.}
	\label{fsdm}
\end{figure}

in NetResSOP, due to the uncertain number of deployed UAVs, the solution dimension should be adjusted iteratively. However, the conventional NSGA-III enforces that the solution dimension remains unchanged to execute SBX and PM. Thus, we introduce a flexible solution dimension mechanism to pad each solution to the largest dimension. More intuitively, Fig. \ref{fsdm}(a) is the original population, while Fig. \ref{fsdm}(b) is the whole population after flexible solution dimension mechanism. Such a flexible solution dimension mechanism pads each solution with the auxiliary decision variables, so that it enlarges the solution to the maximum dimension, while these padded auxiliary decision variables cannot be used for calculating Eqs. (\ref{delta})-(\ref{NetResSOP}).

\subsubsection{Discrete part generation mechanism} there is no method in the conventional NSGA-III to generate the discrete part of the solution. As shown in Fig. \ref{Solution structure}, the discrete part can be divided into the binary matrix and the natural integer. Thus, it may need two different strategies to generate these two parts, respectively. Specifically, the natural integer can be generated as follows:
\begin{equation}
	\label{natural integer}
	\begin{aligned}
		N=\operatorname{randi}(N_{min}, N_{max})
	\end{aligned}
\end{equation}
\noindent where $\operatorname{randi}(N_{min}, N_{max})$ means that generate a random integer between $[N_{min}, N_{max}]$. To enhance the searching efficiency, random search operator and the probabilistic learning operator are introduced. Specifically, these two operators can first utilize similar process to generate an integer which represents the UAV-device pair assignment or the channel allocation, and then convert the integer two a binary matrix. The random search operator is shown in Algorithm \ref{Random_search_operator}, where $A^m_{it}$ is the $m$-th element of $A_{it}$ in $it$-th iteration, which represents that assign the $m$-th UAV-aided D2D pair to the $A^m$-th UAV. For instance, $A^1_{it}=3$ means that the first UAV-aided D2D pair will assign the third UAV. Then, we can generate a $1 \times N$ zero matrix, and assign the third element of the matrix to $1$. At this point, $A_{it}$ will be converted into a binary matrix. Similar representation also applies to $B^n_{it}$, while the discrepancy is that $B^n_{it}$ refers to the channel allocation.

\begin{algorithm}[bp]
	
	\caption{Random search operator}
	\label{Random_search_operator}
	Generate $N$ using Eq. (\ref{natural integer}); \tcc*[f]{Update UAV number}\\ 
	\For{$m$= $1$ to $M$}{ 
		$A^m_{it} = \operatorname{randi}(1, N)$; \tcc*[f]{Update UAV-device pair assignment}\\
	}
	\For{$n$= $1$ to $N+K$}{
		$B^n_{it} = \operatorname{randi}(1, U)$; \tcc*[f]{Update channel allocation}\\
	}
	Execute the flexible solution dimension mechanism;\\
	Transform $A_{it}$ and $B_{it}$ to binary matrices $\mathbb{A}^{M\times N}_{it}$ and $\mathbb{B}^{(N+K)\times U}_{it}$.
	
\end{algorithm}

\par In addition, we provide a probabilistic learning operator to update the discrete part of the solution, which is shown in Algorithm \ref{Probabilistic_learning_operator}, where $\sigma_1$ and $\sigma_2$ are weighting factors. $A_{it}^{\text {Best }}$ is the corresponding part of a random solution in $\mathcal{F}_1$. $\operatorname{randp}(1, N, M)$ is a process to repeatedly utilize the function $\operatorname{randi}$ for $M$ times, which can generate $M$ natural integers, and each of them is between $[1, N]$. Then, the similar process will be imposed to convert $A_{it}$ and $B_{it}$ to the binary matrices $\mathbb{A}^{M\times N}_{it}$ and $\mathbb{B}^{(N+K)\times U}_{it}$.
\begin{algorithm}[tb]
	\caption{Probabilistic learning operator}
	\label{Probabilistic_learning_operator}
	\uIf{$\operatorname{rand}<\sigma_1$}{
		Generate $N$ using Eq. (\ref{natural integer});\\
		$A_{it} = \operatorname{randp}(1, N, M)$, $B_{it} = \operatorname{randp}(1, U, N+K)$;\\
	}
	\uElseIf{$\sigma_1 \leq\operatorname{rand}<\sigma_2$}{
		$A_{it}=A_{it}$, $B_{it}=B_{it}$;\\}
	\Else{
		$A_{it}=A_{it}^{\text {Best }}$, $B_{it}=B_{it}^{\text {Best }}$;\\
	}
	Execute the flexible solution dimension mechanism;\\
	Transform $A_{it}$ and $B_{it}$ to binary matrices $\mathbb{A}^{M\times N}_{it}$ and $\mathbb{B}^{(N+K)\times U}_{it}$.
\end{algorithm}

\subsubsection{UAV number adjustment mechanism}
\par due to the uncertainty of UAV number, the algorithm may need a method to adjust UAV number iteratively. Although it can utilize Eq. (\ref{natural integer}) to adjust UAV number, the search efficiency of the algorithm may be reduced due to the randomness of Eq. (\ref{natural integer}), which may cause the algorithm to get trapped into the local dilemma. Thus, we introduce a UAV number adjustment mechanism, which contains two principles to overcome the shortcomings above, and the principles are given as follows.

\textbf{(a) Reverse walk principle}: if UAV number reaches the bounds, it will execute the reverse walk principle with a step size of $1$. Specifically, when UAV number $N=N_{max}$, it will execute $N=N-1$. Instead, when $N=N_{min}$, it will execute $N=N+1$.

\textbf{(b) Random direction walk principle}: if UAV number is between $(N_{min}, N_{max})$, it can increase or reduce. Thus, we set a increasing probability $p_{in}$ to increase the UAV number, while a reducing probability $1-p_{in}$ is used to reduce the UAV number. The increasing or reducing step size is also set as $1$. Then, assume $\operatorname{rand}$ is a random number that generated from $(0, 1)$, and the main steps of the UAV number adjustment mechanism are presented in Algorithm \ref{UAVrelayadjustment}.

\subsection{The methods to deal with constraints}
\par As can be seen, $C1 - C9$ are strong constraints that limit the maximum and minimum bounds and the domains of the decision variables, while $C10$ is a weak constraint to limit the flight times of UAVs. For this purpose, $C1 - C9$ will be coped in the initialization and iteration process of the multi-objective evolutionary algorithms to ensure their legalities. Moreover, $C10$ will be regarded as a penalty function. Then, the detailed methods to deal with different constraints are as follows. \textbf{(a)} The method to deal with $C1 - C5$: for continuous solutions $x_n$, $y_n$, $z_n$, $P_n$, and $V_n$, we add a step to judge whether the solution satisfy the constraints in $C1 - C5$. Once the solution does not satisfy the constraint, the original solution generation method of multi-objective evolutionary algorithm is used to correct it. For example, if $x_n > L_{max}$, $x_n = L_{min}+ \operatorname{rand} \times (L_{max}-L_{min})$. \textbf{(b)} The method to deal with $C6 - C9$: these constraints limit the domains of the decision variables which represent the UAV-device pair assignment, the channel allocation and UAV number. Thus, we can first generate the corresponding UAV number and channel number, and then convert them into the binary variables for computation. For instance, if the first UAV-aided D2D pair selects the third UAV, ``3'' is generated. Then, we can generate a $1 \times N$ zero matrix, and assign the third element of the matrix to $1$. When updating the solutions, we only update ``3'' using $\operatorname{randi}(1, N)$. Similar operators will be executed for multiple times to update the UAV-device pair assignment and the channel allocation. \textbf{(c)} The method to deal with $C10$: $C10$ is a weak constraint, and hence it is treated as a penalty function. Once $C10$ is not satisfied, the values of the three optimization objectives will be added with a very large number to reduce the quality of the solution, respectively, so that the algorithm discards these solutions with poor quality during the iteration process. In this work, we choose $10^7$, $8$, and $10^6$ as three penalty items for the three optimization objectives, respectively. If $C10$ is not satisfied, $-f_1 = -f_1 + 10^7$, $f_2 = f_2 + 8$, $f_3 = f_3 +10^6$.

	\begin{algorithm}[t]
		\caption{UAV number adjustment mechanism}
		\label{UAVrelayadjustment}
		\uIf{$N=N_{max}$}{
			$N=N_{max}-1$;\\
		}
		\uElseIf{$N=N_{min}$}{
			$N=N_{min}+1$;\\}
		\Else{\eIf{$\operatorname{rand}<p_{in}$}{$N=N+1$;\\}{$N=N-1$;}
		}
	\end{algorithm}
\subsection{Algorithm analysis}
\subsubsection{Complexity} referred to \cite{DBLP:journals/tec/DebJ14}, the complexity of the conventional NSGA-III is the larger of $\mathcal{O}(Pop^2\cdot \log_2^{o-2} Pop)$ and $\mathcal{O}(Pop^2\cdot o)$, where $o$ is the number of objective functions. The improvement of the NSGA-III-FDU is to embed a flexible solution dimension mechanism into the initialization of the conventional NSGA-III, introduce a discrete part generation mechanism to update the discrete part of the solution, and introduce a UAV number adjustment mechanism to update the UAV number. First, the flexible solution dimension mechanism cannot influence the complexity, since it only improves the dimensions of some solutions to the maximum size. Second, according to Algorithm \ref{NSGA-III-FDU}, the discrete part generation mechanism and UAV number adjustment mechanism only increase $|\mathcal{Q}^\prime_{it}|=Pop$ computations of the objective functions. Thus, the complexity of the NSGA-III-FDU is the larger of $\mathcal{O}(2Pop^2\cdot \log_2^{o-2} Pop)$ and $\mathcal{O}(2Pop^2\cdot o)$. When $Pop$ is sufficiently large, the complexity of the NSGA-III-FDU is the larger of $\mathcal{O}(Pop^2\cdot \log_2^{o-2} Pop)$ and $\mathcal{O}(Pop^2\cdot o)$, which is the same as the conventional NSGA-III.

\subsubsection{Convergence} the proposed NSGA-III-FDU is a variant of the conventional NSGA-II, which is still an evolutionary algorithm. The evolutionary algorithms will converge on unimodal and multimodal functions\cite{DBLP:conf/cec/Zhao08}. Moreover, the authors in \cite{DBLP:journals/isci/ChenH21}  analyze the average convergence rate on continuous optimization, which represents a normalized geometric mean of the reduction ratio of the fitness difference per iteration, while the authors in \cite{DBLP:journals/tec/HeL16} analyze average convergence rate on discrete optimization. Thus, NSGA-III-FDU converges.
\subsubsection{Drawbacks} although NSGA-III-FDU is suitable for solving NetResSOP, it is still based on the framework of evolutionary algorithms. Thus, it has the common drawbacks of evolutionary algorithms as follows. 1) These algorithms cannot guarantee to obtain the global optimal solution when the search space is complex. 2) These algorithms have a certain dependence on the initial population and need to be improved by combining some heuristic information. 3) These algorithms are difficult to analyze theoretically, and the parameters of the algorithms may influence the results. It is basically impossible to definitely state that a certain type of algorithm is most suitable for a certain type of problem. Nevertheless, evolutionary algorithms are still effective ways to solve NP-hard problems and MIPPs \cite{zhao2023self} \cite{DBLP:journals/tec/HuynhCSR18}, and they are practical for the real scenarios. In the real scenarios, large-scale optimization problems are often faced, and it will take an inestimable time to find the global optimal solution. Thus, a feasible strategy is to use evolutionary algorithms, which can obtain an acceptable solution in polynomial time and can deal with multiple constraints.
\section{Simulation results}
\label{Simulation results}
\par In this section, simulations results are given based on Matlab to illustrate the performance of the proposed NSGA-III-FDU.
\subsection{Parameter setups}
\par The horizontal area is set as $400$ m $\times$ $400$ m, and the maximum and the minimum altitudes of UAVs are
set as $500$ m and $200$ m, respectively, and all UAVs are initially at $(0, 0, 200)$. The maximum and the minimum flight velocities of UAVs are set as $16$ m/s and $6$ m/s, respectively. The preset time threshold $T_{th}$ is set as $12$ s. Other parameters about UAV energy consumption in Eqs. (\ref{UAV-2D-Power}) and (\ref{UAV-3D-Power}) refer to \cite{DBLP:journals/tcom/SunLWWSL22}. Moreover, the path loss at the reference distance $\beta_0$ and the path loss exponent $\alpha$ are set as $-60$ dbm and $2$, respectively. The bandwidths of all channels are set as $W=1$ MHz, the center frequency of carrier is set as $f_c=2$ GHz, and the white Gaussian noise power $\sigma^2$ is set as $-174$ dBm/Hz. In addition, the transmission powers of all SWDs are set as $0.01$ W, while the maximum and the minimum transmission powers of UAVs are set as $1$ W and $0.1$ W, respectively. The activities of direct D2D pairs are set as $Pr_k=0.6$. Besides, the parameters of channel conditions are set as $a=9.61$, $b=0.16$, $\eta_{LoS}=1$ and $\eta_{NLoS}=20$ \cite{DBLP:journals/twc/LuWCCFL18}, which are the standard parameters in urban environments. The population size and the maximum iteration of these algorithms are set as $20$ and $200$, respectively. $\sigma_1$ and $\sigma_2$ are set as $0.2$ and $0.6$, respectively. Finally, we consider two different scales of D2D networks, which can be recorded as the small scale D2D network (Scale $1$) and the large scale D2D network (Scale $2$), respectively, and the D2D network parameters of these two scales are given in Table \ref{Network parameters}. 

\begin{table}[tb]
	\setlength{\abovedisplayskip}{1pt}
	\setlength{\belowdisplayskip}{1pt}
	\setlength{\abovecaptionskip}{5pt}
	\tiny
	\begin{center}
		\caption{Parameters for different scale D2D networks}
		\setlength{\tabcolsep}{2mm}
		{\begin{tabular}{ccc}\toprule  
				D2D network parameters &Scale 1  &{Scale 2}  \\
				\midrule
				$N_{max}$  & $8$  &{$16$} \\\cmidrule(lr){2-3}
				$N_{min}$  & $4$  &{$8$} \\\cmidrule(lr){2-3}   
				$U$  & $3$  &{$7$} \\\cmidrule(lr){2-3}
				$M$  & $10$  &{$100$} \\\cmidrule(lr){2-3}
				$K$  & $3$  &{$6$} \\\bottomrule	    	
		\end{tabular}}
		\label{Network parameters}
	\end{center}
\end{table}
\subsection{Benchmarks}
\par Several benchmarks are adopted, and the details are as follows.
\begin{itemize}
	\item \textbf{Proposed NSGA-III-FDU}: this is our proposed approach described in Algorithm \ref{NSGA-III-FDU} in Section \ref{Algorithm}.
	\item \textbf{Uniform deployment (UD)}: the number of deployed UAVs is set as $(N_{max}+N_{min})/2$, and the UAVs are uniformly deployed at the horizontal area, while the heights of all UAVs are set as $(Z_{max}+Z_{min})/2$. Moreover, the transmit powers of all UAVs are set as the maximum value to obtain the best communication performance. In addition, the flight velocities, channel allocation and UAV-device pair assignment are random. Such a UD strategy is reasonable and common while considering the UAV position deployment problem \cite{10012331}.
	\item \textbf{Random deployment (RD)}: all decision variables are randomly generated within the domain.
	\item \textbf{Linear weighting method}: we convert the three optimization objectives into one optimization objective by using a linear weighting method. Then, a single-objective evolutionary algorithm called mountain gazelle optimizer (MGO) \cite{abdollahzadeh2022mountain}, which is a recently proposed algorithm with good performance, is used to solve it. 
	\item \textbf{State-of-the-art multi-objective comparison algorithms}: several multi-objective optimization algorithms, which are multi-objective stochastic paint optimizer (MOSPO) \cite{khodadadi2022multi}, multi-objective evolutionary algorithm based on decomposition (MOEA/D) \cite{DBLP:journals/tec/ZhangL07}, multi-objective particle swarm optimization (MOPSO) \cite{DBLP:journals/isci/TripathiBP07}, NSGA-II \cite{DBLP:journals/tcyb/ShenWW22}, non-dominated sorting genetic algorithm-III (NSGA-III) \cite{DBLP:journals/tec/DebJ14}, and improved multi-objective salp swarm algorithm (IMSSA) \cite{sun2023uav} are employed as comparison algorithms.
\end{itemize}

\subsection{Optimization results}
\par In this part, we give the visualization of optimization results, verify the convergence and the optimality, and evaluate the performance of the algorithm under different alternative strategies.

\begin{table}[t]
	\setlength{\abovedisplayskip}{1pt}
	\setlength{\belowdisplayskip}{1pt}
	\setlength{\abovecaptionskip}{1pt}
	\tiny
	\tabcolsep=1.5mm
	\begin{center}
		\caption{Numerical optimization results obtained by various comparison methods}
		\begin{tabular}{lllllll}
			\toprule
			\multirow{2}{*}{Benchmarks} & \multicolumn{3}{c}{Scale 1} &\multicolumn{3}{c}{Scale 2}        \\ \cmidrule(l){2-7} 
			& $f_1$ [bps] & $f_2$ & $f_3$ [J]& $f_1$ [bps]& $f_2$ & $f_3$ [J]          \\ \midrule
			UD & $-9.88 \times 10^6$ &$14$ & $1.00 \times 10^6$ & $-7.52 \times 10^6$&$20$ & $1.00 \times 10^6$\\
			RD & $-9.58 \times 10^6$ &$13$ & $1.00 \times 10^6$ & $-7.83 \times 10^6$&$24$ & $1.00 \times 10^6$\\ 
			MGO &$9.01 \times 10^5$ &$4$ & $6.69 \times 10^3$ & $-7.55 \times 10^6$&$16$ & $1.00 \times 10^6$\\ 
			MOSPO &$1.55 \times 10^6$ &$\bm{4}$ & $4.09 \times 10^3$ & $2.30 \times 10^6$ & $\bm{8}$ & $6.76 \times 10^3$\\ 
			MOPSO &$1.35 \times 10^6$ &$\bm{4}$ & $4.94 \times 10^3$ & $1.91 \times 10^6$&$\bm{8}$ & $7.31 \times 10^3$\\ 
			MOEA/D &$1.20 \times 10^6$ &$\bm{4}$ & $5.34 \times 10^3$ & $-7.49 \times 10^6$&$16$ & $1.00 \times 10^6$\\
			NSGA-II &$1.27 \times 10^6$ &$\bm{4}$ & $2.08 \times 10^3$ & $2.70 \times 10^6$&$\bm{8}$ & $3.34 \times 10^3$\\  
			NSGA-III &$8.42 \times 10^5$ &$\bm{4}$ & $2.29 \times 10^3$ & $2.19 \times 10^6$&$\bm{8}$ & $3.76 \times 10^3$\\  
			IMSSA &$1.44 \times 10^6$ &$\bm{4}$ & $3.37 \times 10^3$ & $2.69 \times 10^6$&$\bm{8}$ & $6.49 \times 10^3$\\ 
			NSGA-III-FDU &$\bm{2.01 \times 10^6}$ &$\bm{4}$ & $\bm{1.78 \times 10^3}$ & $\bm{5.07 \times 10^6}$&$\bm{8}$ & $\bm{2.74 \times 10^3}$\\       
			\bottomrule
		\end{tabular}
		\label{table:strategies_algorithms}
	\end{center}
\end{table}

\begin{figure}[t]
	\centering{\includegraphics[width=3.5in]{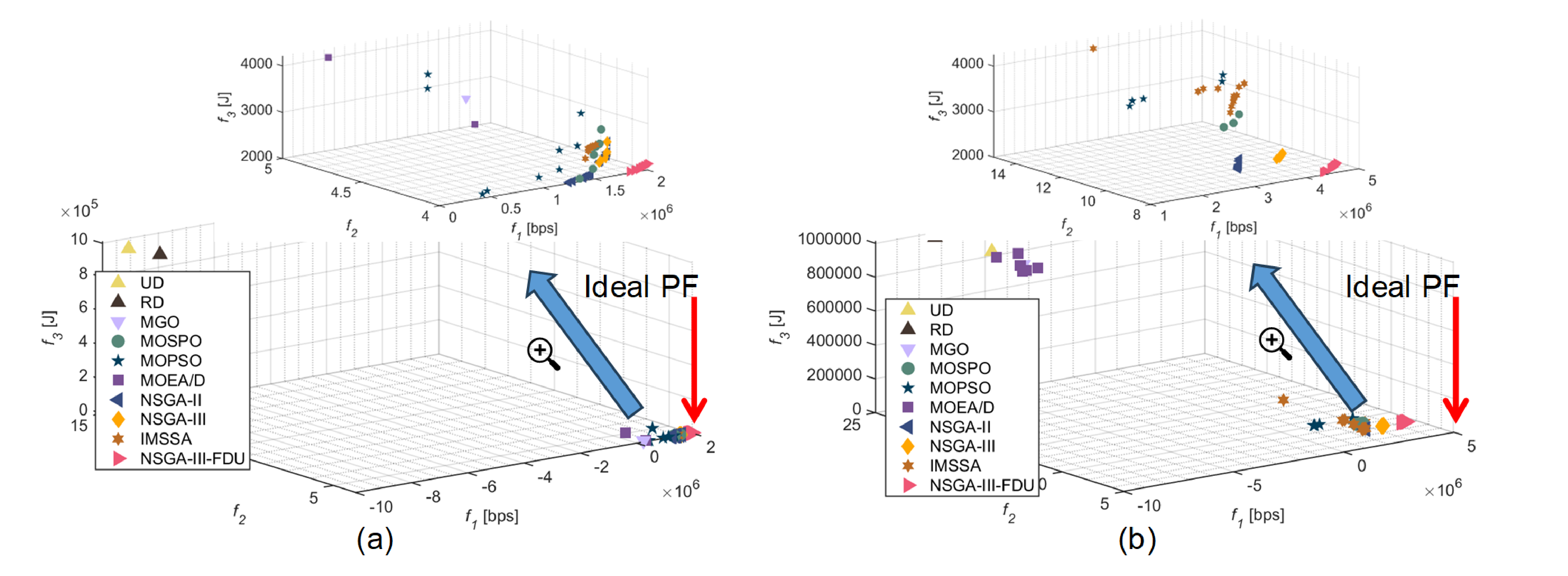}}
	\caption{Solution distribution obtained by different benchmarks. (a) Scale 1. (b) Scale 2.}
	\label{Pareto_results}
\end{figure}

\subsubsection{Visualization of optimization results} 
\begin{figure}[t]
	\vspace{-0.4cm}
	\centering{\includegraphics[width=3.5in]{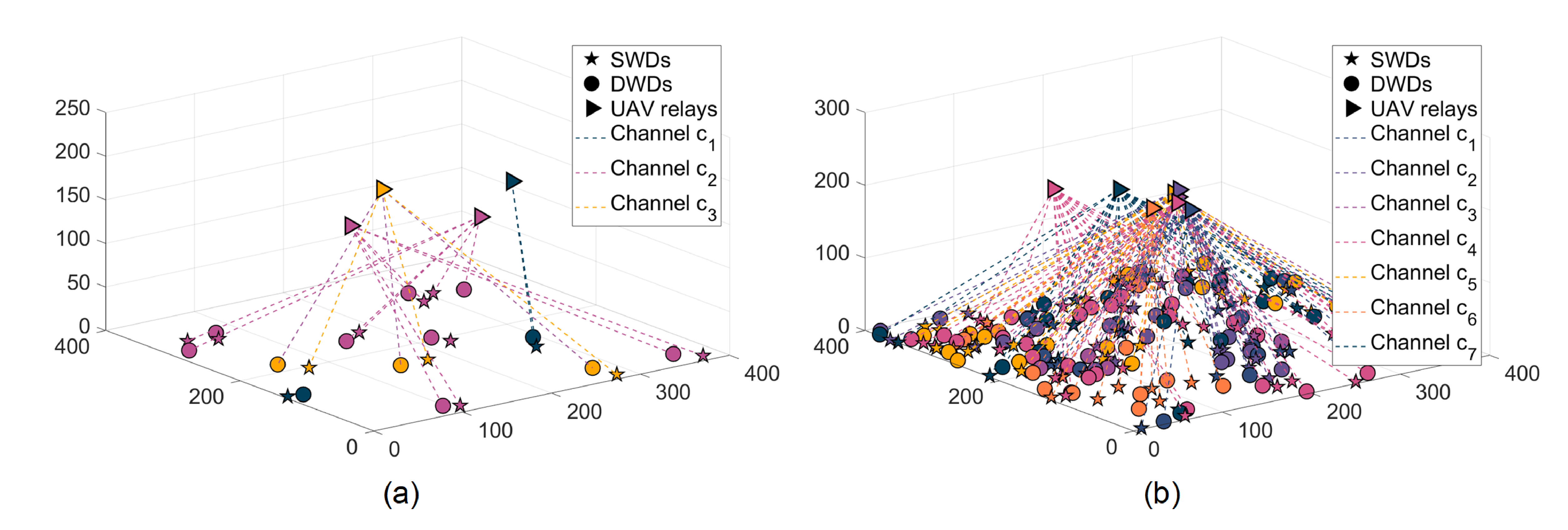}}
	\caption{Optimization results of UAV position deployment, channel allocation, UAV-device pair assignment, and UAV number obtained by the proposed NSGA-III-FDU for different scales of D2D networks.  (a) Scale 1, $N=4$. (b) Scale 2, $N=8$.}
	\label{Display_connection}
\end{figure}

\begin{figure}[t]
	\centering{\includegraphics[width=3.5in]{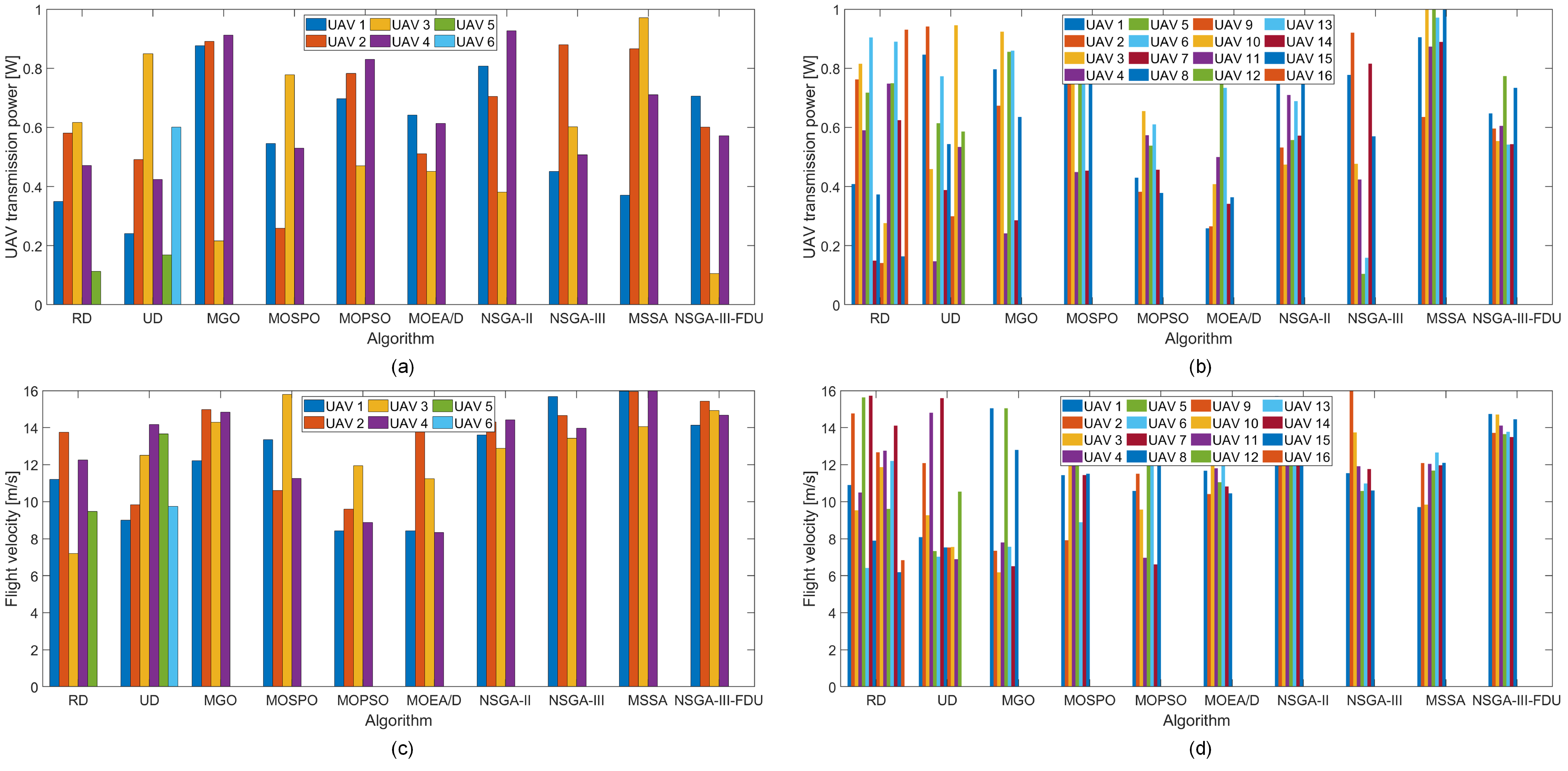}}
	\caption{UAV transmission powers and UAV flight velocities obtained by the proposed NSGA-III-FDU and other benchmarks.  (a) Transmission power, Scale 1, $N=4$. (b) Transmission power, Scale 2, $N=8$. (c) Velocities, Scale 1, $N=4$. (d) Velocities, Scale 2, $N=8$.}
	\label{Power_and velocity}
\end{figure}

Table \ref{table:strategies_algorithms} shows the numerical optimization results obtained by various comparison methods. As can be seen, the negative values on $f_1$ mean that the results cannot satisfy the $C10$, and hence a penalty will be imposed on the three optimization objectives, resulting in $f_2$ exceeding the boundary conditions and $f_3$ being extremely large. In Scale 1, UD and RD strategies cannot satisfy the constraints, while more comparison algorithms, such as MGO and MOEA/D, cannot satisfy the constraints in Scale 2. Moreover, the proposed NSGA-III-FDU can obtain the best performance in both Scales 1 and 2. Then, Figs. \ref{Pareto_results}(a) and \ref{Pareto_results}(b) illustrate the corresponding solution distribution obtained by different benchmarks. Specifically, the three dimensions in Fig. \ref{Pareto_results} represent the three optimization objectives of Eq. (\ref{NetResSOP}a), and we can find the superiority relationship of these algorithms from the figure intuitively. Obviously, the solutions obtained by NSGA-III-FDU are much closer to the direction of the ideal PF. These results indicate that the proposed NSGA-III-FDU is more suitable for solving the formulated NetResSOP. The reasons may be the discrete part generation mechanism increases the search efficiency, and the UAV number adjustment mechanism can utilize the greedy strategy to select the suitable UAV number adaptively, which can further enhance the performance of the NSGA-III-FDU. Intuitively, Figs. \ref{Display_connection}(a) and \ref{Display_connection}(b) show that the UAV position deployment, channel allocation, UAV-device pair assignment, and UAV number obtained by the proposed NSGA-III-FDU under Scales 1 and 2, respectively, while Fig. \ref{Power_and velocity} shows the transmission powers and velocities of the UAVs, when using the minimum number of UAVs.

\subsubsection{Convergence and optimality verification}
\begin{figure}[t]
	\centering{\includegraphics[width=3.5in]{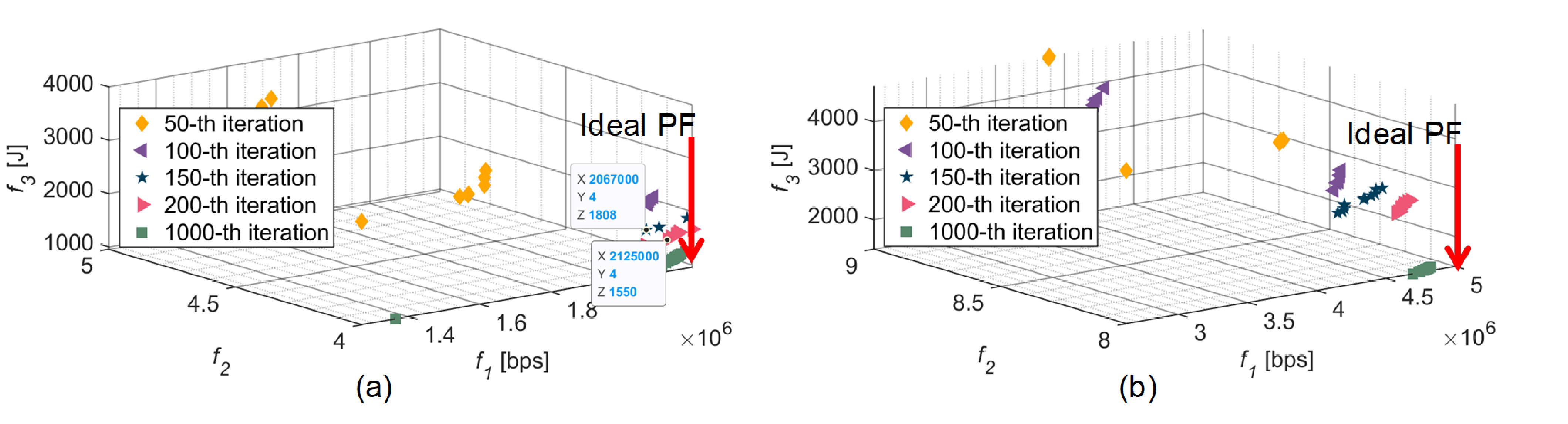}}
	\caption{Advanced progress of PF to ideal PF and gap with the approximate-optimal PF obtained by NSGA-III-FDU. (a) Scale 1. (b) Scale 2.}
	\label{Convergence}
\end{figure}

\par since the formulated NetResSOP with three optimization objectives is an MOP with trade-offs, it is difficult to find the optimal value for each optimization objective simultaneously. Thus, it is reasonable to use the advanced process of PF to verify the convergence of the algorithm by introducing the concept of Pareto dominance. Specifically, Fig. \ref{Convergence}(a) shows the advanced process of PF in different iterations for Scale 1, while Fig. \ref{Convergence}(b) shows the corresponding results for Scale 2. It can be seen from Fig. \ref{Convergence}(a) that the obtained PF on $200$-th iteration is better than that on $150$-th iteration, which means that the obtained solution qualities are getting better with more iterations, while the improvement in solution qualities decreases gradually. Thus, the algorithm gradually converges. Moreover, the similar result can also be found in  Fig. \ref{Convergence}(b), while the magnitude of the advance on PF from $50$-th iteration to $100$-th iteration is less obvious than that in Scale 1. The reason may be that the solution dimension of Scale 2 is larger and the solution space of Scale 2 is more complex, which means that the algorithm needs more searches to find a better PF.

\par Note that finding the approximate-optimal solution for the problem may be more reasonable, which is also a common way for the MOPs in wireless systems and experimentations \cite{DBLP:journals/twc/MehariPCDVPJMDM16}, since it is difficult to find an optimal solution that can simultaneously make all objectives be optimal. To analyze the approximate gap in a feasible way, we first set the number of iterations of the proposed algorithm to $1000$, which is a much larger value than the normal simulations. Then, the obtained optimization result is regarded as the approximate-optimal PF, and we compare the gap between the result of the normal simulations and the approximate-optimal PF. Fig. \ref{Convergence}(a) shows that the PF obtained by $200$-th iteration is close to the approximate-optimal PF for Scale 1. For Scale 2, the gap from $200$-th iteration to approximate-optimal PF is also not very large.

\par In addition, we further explain the Pareto dominance of the solutions according to the results. Specifically, Fig. \ref{Convergence}(a) shows the advanced process of PF in Scale 1 during the iteration process, wherein each dimension corresponds to one optimization objective of the formulated NetResSOP. Moreover, the triangles with pink color represent the PF on $200$-th iteration and the pentagrams with dark-blue color represent the PF on $150$-th iteration. Then, we select two solutions, whose values of three optimization objectives are $\Lambda_1= (2067000, 4, 1808)$ and $\Lambda_2= (2125000, 4, 1550)$, and the corresponding values of Eq. (\ref{NetResSOP}a) are $F(\mathbf{X}_1) = (-2067000, 4, 1808)$ and $F(\mathbf{X}_2) = (-2125000, 4, 1550)$. Obviously, $F(\mathbf{X}_2)\prec F(\mathbf{X}_1)$, which means that $\Lambda_2$ will obtain greater D2D network capacity with less energy consumption when deploying the same number of UAVs compared to $\Lambda_1$. Thus, during the iteration process, a great number of dominated solutions, such as $\Lambda_1$, will be discarded, while the non-dominated solutions, such as $\Lambda_2$, will be remained.

\subsubsection{Performance under different alternative strategies}
\par due to the nature of multi-objective evolutionary algorithms,  decision-makers can obtain a set of solutions with Pareto dominance, and they need to select one from the solution set according to the real requirements \cite{10012331}. For example, when number of available UAVs is less, we must utilize the limited UAV resources to achieve a greater D2D network capacity, while reducing the energy consumption of UAVs. Instead, when UAV number is not the primary consideration, we can increase the D2D network capacity as soon as possible, while reducing the number of deployed UAVs and the energy consumption of UAVs. Additionally, when UAVs perform special tasks, it is necessary to choose a UAV of smaller size, which means that the on-board energy of each UAV is less. Under the circumstances, we need to optimize energy consumption of each UAV, while maximizing the D2D network capacity and minimizing the number of deployed UAVs. Thus, we design three alternative strategies for decision-makers, which are MaxNetCap, MinUAV and MinAveEnergy. 
\par Tables \ref{numerical results of USOP by MaxNetCap}, \ref{numerical results of USOP by MinUAV} and \ref{numerical results of USOP by MinAveEnergy} show the statistical results obtained by NSGA-III-FDU and other comparison approaches by using MaxNetCap, MinUAV and MinAveEnergy, respectively, wherein ``Mean'', ``Std.'', ``Max'' and ``Min'' represent the mean value, standard deviation, maximum value and minimum value of $30$ independent trials. Similar to Table \ref{table:strategies_algorithms}, the negative values on $f_1$ means that the solutions cannot satisfy $C10$. As can be seen, for any strategy of MaxNetCap, MinUAV and MinAveEnergy, the proposed NSGA-III-FDU obtains the best performance under both Scales 1 and 2, and it can satisfy $C10$ for all $30$ independent trials, which means that NSGA-III-FDU is better and more stable than other comparison algorithms.

\begin{table*}[t]
	\setlength{\abovedisplayskip}{1pt}
	\setlength{\belowdisplayskip}{0pt}
	\setlength{\abovecaptionskip}{1pt}
	\tiny
	\tabcolsep=0.15cm
	\begin{center}
		\caption{Statistical results of MaxNetCap obtained by NSGA-III-FDU and other multi-objective comparison approaches for NetResSOP}
		{\begin{tabular}{cccccccccc}\toprule
				&&\textbf{MOSPO} &\textbf{MOPSO} &\textbf {MOEA/D} &\textbf{NSGA-II} &\textbf{NSGA-III} &\textbf{IMSSA}&\textbf{NSGA-III-FDU}&\textbf{Improvement}\\\cmidrule(lr){3-10}&&\multicolumn{7}{c}{\textbf {Scale 1}}\\\cmidrule(lr){1-10}
				\multirow{4}{*}{$f_1$}
				&Mean&\multicolumn{1}{|c}{$1.25\times 10^6$}&$9.63\times 10^5$&$1.09\times 10^6$&$1.65\times 10^6$&$1.68\times 10^6$&$1.63\times 10^6$&$\bm{2.09\times10^6}$&$24.40\%$\\
				&Std.&\multicolumn{1}{|c}{$3.14\times 10^5$}&$4.29\times 10^5$&$4.34\times 10^5$&$3.55\times 10^5$&$4.00\times 10^5$&$1.63\times 10^5$&$\bm{1.08\times 10^5}$&\\
				&Max&\multicolumn{1}{|c}{$1.89\times 10^6$}&$1.62\times 10^6$&$1.44\times 10^6$&$2.25\times 10^6$&$\bm{2.59\times 10^6}$&$2.04\times 10^6$&$2.33\times 10^6$&\\
				&Min&\multicolumn{1}{|c}{$5.82\times 10^5$}&$1.35\times 10^6$&$-9.22\times 10^5$&$6.68\times 10^5$&$5.02\times 10^5$&$1.37\times 10^6$&$\bm{1.84\times 10^6}$&\\\cmidrule(lr){1-10}
				\multirow{4}{*}{$f_2$}
				&Mean&\multicolumn{1}{|c}{5.50}&4.73&7.20&4.60&4.26&\textbf{4.03}&4.23&-4.96\%\\
				&Std.&\multicolumn{1}{|c}{1.25}&0.94&3.98&1.10&0.44&\textbf{0.18}&0.43&\\
				&Max&\multicolumn{1}{|c}{8.00} &7.00&8.00&15.00&\textbf{5.00}&\textbf{5.00}&\textbf{5.00}&\\
				&Min&\multicolumn{1}{|c}{\textbf{4.00}}&\textbf{4.00}&\textbf{4.00}&\textbf{4.00}&\textbf{4.00}&\textbf{4.00}&\textbf{4.00}&\\\cmidrule(lr){1-10}
				\multirow{4}{*}{$f_3$}
				&Mean&\multicolumn{1}{|c}{$5.14\times 10^3$}&$6.28\times 10^3$&$2.73\times 10^5$&$2.64\times 10^3$&$2.47\times 10^3$&$4.25\times 10^3$&$\bm{2.34\times 10^3}$&$5.26\%$\\
				&Std.&\multicolumn{1}{|c}{$1.10\times 10^3$}&$\bm{1.16\times 10^2}$&$4.49\times 10^5$&$6.31\times 10^2$&$7.60\times 10^2$&$9.15\times 10^2$&$3.93\times10^2$\\
				&Max&\multicolumn{1}{|c}{$7.30\times 10^3$}&$8.94\times 10^3$&$1.00\times 10^6$&$4.14\times 10^3$&$4.17\times 10^3$&$5.95\times 10^3$&$\bm{3.44\times 10^3}$\\
				&Min&\multicolumn{1}{|c}{$3.31\times 10^3$}&$4.39\times 10^3$&$4.20\times 10^3$&$1.73\times 10^3$&$\bm{1.17\times 10^3}$&$2.48\times 10^3$&$1.73\times10^3$&\\\cmidrule(lr){1-10}
				&&\multicolumn{7}{c}{\textbf {Scale 2}}\\\cmidrule(lr){1-10}
				\multirow{4}{*}{$f_1$}
				&Mean&\multicolumn{1}{|c}{$1.58\times 10^6$}&$-8.65\times 10^5$&$-6.86\times 10^6$&$3.43\times 10^6$&$2.98\times 10^6$&$3.53\times 10^6$&$\bm{4.96\times 10^6}$&$40.50\%$\\
				&Std.&\multicolumn{1}{|c}{$3.05\times 10^6$}&$4.47\times 10^6$&$3.40\times 10^5$&$1.08\times 10^6$&$2.46\times 10^6$&$3.45\times 10^5$&$\bm{3.11\times 10^5}$\\
				&Max&\multicolumn{1}{|c}{$3.50\times 10^6$}&$3.59\times 10^6$&$-6.09\times 10^6$&$4.94\times 10^6$&$4.73\times 10^6$&$4.43\times 10^6$&$\bm{5.59\times 10^6}$\\
				&Min&\multicolumn{1}{|c}{$-7.82\times 10^6$}&$-6.89\times 10^6$&$-7.39\times 10^5$&$1.14\times 10^6$&$-6.02\times 10^6$&$3.01\times 10^6$&$\bm{4.01\times 10^6}$\\\cmidrule(lr){1-10}
				\multirow{4}{*}{$f_2$}
				&Mean&\multicolumn{1}{|c}{11.30}&14.16&21.26&9.06&9.73&10.33&\textbf{8.96}&1.10\%\\
				&Std.&\multicolumn{1}{|c}{3.90}&6.16&2.39&1.55&2.43&2.60&\textbf{1.40}\\
				&Max&\multicolumn{1}{|c}{22.00}&24.00&24.00&14.00&17.00&16.00&\textbf{15.00}\\
				&Min&\multicolumn{1}{|c}{\textbf{8.00}}&\textbf{8.00}&18.00&\textbf{8.00}&\textbf{8.00}&\textbf{8.00}&\textbf{8.00}\\\cmidrule(lr){1-10}
				\multirow{4}{*}{$f_3$}
				&Mean&\multicolumn{1}{|c}{$1.06\times 10^5$}&$4.06\times 10^5$&$1.00\times 10^6$&$3.08\times 10^3$&$6.97\times 10^4$&$5.86\times 10^3$&$\bm{2.91\times 10^3}$&$5.51\%$\\
				&Std.&\multicolumn{1}{|c}{$3.05\times 10^5$}&$\bm{4.98\times 10^5}$&$7.07\times 10^2$&$4.45\times 10^2$&{$2.53\times 10^5$}&$7.32\times 10^2$&$5.60\times 10^2$\\
				&Max&\multicolumn{1}{|c}{$1.00\times 10^6$}&$1.00\times 10^6$&$1.00\times 10^6$&$\bm{4.06\times 10^3}$&{$1.00\times 10^6$}&$6.74\times 10^3$&$5.13\times 10^3$\\
				&Min&\multicolumn{1}{|c}{$4.82\times 10^3$}&$5.36\times 10^3$&$1.00\times 10^6$&$2.53\times 10^3$&$2.45\times 10^3$&$3.96\times 10^3$&$\bm{1.95\times 10^3}$\\\bottomrule
		\end{tabular}}
		\label{numerical results of USOP by MaxNetCap}
	\end{center}
\end{table*}
\begin{table*}[t]
	\setlength{\abovedisplayskip}{1pt}
	\setlength{\belowdisplayskip}{1pt}
	\setlength{\abovecaptionskip}{1pt}
	\vspace{-0.6cm}
	\tiny
	\tabcolsep=0.15cm
	\begin{center}
		\caption{Statistical results of MinUAV obtained by NSGA-III-FDU and other multi-objective comparison approaches for NetResSOP}

		{\begin{tabular}{cccccccccc}\toprule
				&&\textbf{MOSPO} &\textbf{MOPSO} &\textbf {MOEA/D} &\textbf{NSGA-II} &\textbf{NSGA-III} &\textbf{IMSSA} &\textbf{NSGA-III-FDU}&\textbf{Improvement}\\\cmidrule(lr){3-10}&&\multicolumn{7}{c}{\textbf {Scale 1}}\\\cmidrule(lr){1-10}
				\multirow{4}{*}{$f_1$}
				&Mean&\multicolumn{1}{|c}{$7.60\times 10^5$}&$7.95\times 10^5$&$-1.91\times 10^6$&$1.45\times 10^6$&$1.53\times 10^6$&$1.48\times 10^6$&$\bm{2.01\times 10^6}$&$31.37\%$\\
				&Std.&\multicolumn{1}{|c}{$3.95\times 10^5$}&$3.65\times 10^5$&$4.33\times 10^6$&$5.41\times 10^5$&$4.62\times 10^5$&$3.54\times 10^5$&$\bm{1.15\times 10^5}$\\
				&Max&\multicolumn{1}{|c}{$1.59\times 10^5$}&$1.31\times 10^6$&$1.42\times 10^6$&$2.25\times 10^6$&$\bm{2.59\times 10^6}$&$2.04\times 10^6$&$2.25\times 10^6$\\
				&Min&\multicolumn{1}{|c}{$2.39\times 10^5$}&$1.35\times 10^5$&$-9.29\times 10^6$&$2.25\times 10^5$&$4.69\times 10^5$&$1.81\times 10^5$&$\bm{1.77\times 10^6}$\\\cmidrule(lr){1-10}
				\multirow{4}{*}{$f_2$}
				&Mean&\multicolumn{1}{|c}{4.23}&4.33&6.53&4.03&4.06&\textbf{4.00}&\textbf{4.00}&0.00\%\\
				&Std.&\multicolumn{1}{|c}{0.62}&0.75&3.64&0.18&0.25&\textbf{0.00}&\textbf{0.00}\\
				&Max&\multicolumn{1}{|c}{7.00}&7.00&13.00&5.00&6.00&\textbf{4.00}&\textbf{4.00}\\
				&Min&\multicolumn{1}{|c}{\textbf{4.00}}&\textbf{4.00}&\textbf{4.00}&\textbf{4.00}&\textbf{4.00}&\textbf{4.00}&\textbf{4.00}\\\cmidrule(lr){1-10}
				\multirow{4}{*}{$f_3$}
				&Mean&\multicolumn{1}{|c}{$5.47\times 10^3$}&$5.61\times 10^3$&$2.73\times 10^5$&$2.39\times 10^3$&$2.19\times 10^3$&{$3.90\times 10^3$}&$\bm{2.08\times 10^3}$&$5.02\%$\\
				&Std.&\multicolumn{1}{|c}{$1.98\times 10^3$}&$1.56\times 10^3$&$4.49\times 10^5$&$6.66\times 10^2$&$6.26\times 10^2$&{$6.92\times 10^2$}&$\bm{3.06\times 10^2}$\\
				&Max&\multicolumn{1}{|c}{$1.01\times 10^4$}&$8.94\times 10^3$&$1.00\times 10^6$&$5.31\times 10^3$&$4.17\times 10^3$&{$5.45\times 10^3$}&$\bm{2.57\times 10^3}$\\
				&Min&\multicolumn{1}{|c}{$1.96\times 10^3$}&$2.67\times 10^3$&$4.20\times 10^3$&$1.73\times 10^3$&$\bm{1.15\times 10^3}$&{$2.48\times 10^3$}&$1.45\times 10^3$\\\cmidrule(lr){1-10}
				&&\multicolumn{7}{c}{\textbf {Scale 2}}\\\cmidrule(lr){1-10}
				\multirow{4}{*}{$f_1$}
				&Mean&\multicolumn{1}{|c}{$1.19\times 10^6$}&$-1.83\times 10^6$&$-7.23\times 10^6$&$3.15\times 10^6$&$2.63\times 10^6$&$3.02\times 10^6$&$\bm{4.33\times 10^6}$&$ 37.46\%$\\
				&Std.&\multicolumn{1}{|c}{$3.25\times 10^6$}&$4.88\times 10^6$&$\bm{3.83\times 10^5}$&$1.03\times 10^6$&$2.42\times 10^6$&$5.37\times 10^5$&$8.07\times 10^5$\\
				&Max&\multicolumn{1}{|c}{$3.50\times 10^6$}&$3.18\times 10^6$&$-6.46\times 10^6$&$4.91\times 10^6$&$4.69\times 10^6$&$4.37\times 10^6$&$\bm{5.32\times 10^6}$\\
				&Min&\multicolumn{1}{|c}{$-8.57\times 10^6$}&$-8.71\times 10^6$&$-7.99\times 10^6$&$1.13\times 10^6$&$-6.03\times 10^6$&$\bm{2.20\times 10^6}$&$1.98\times 10^6$\\\cmidrule(lr){1-10}
				\multirow{4}{*}{$f_2$}
				&Mean&\multicolumn{1}{|c}{10.26}&11.30&18.83&8.46&8.83&\textbf{8.00}&\textbf{8.00}&0.00\%\\
				&Std.&\multicolumn{1}{|c}{3.39}&3.92&2.22&1.22&2.15&\textbf{0.00}&\textbf{0.00}\\
				&Max&\multicolumn{1}{|c}{21.00}&16.00&24.00&14.00&17.00&\textbf{8.00}&\textbf{8.00}\\
				&Min&\multicolumn{1}{|c}{\textbf{8.00}}&\textbf{8.00}&16.00&\textbf{8.00}&\textbf{8.00}&\textbf{8.00}&\textbf{8.00}\\\cmidrule(lr){1-10}
				\multirow{4}{*}{$f_3$}
				&Mean&\multicolumn{1}{|c}{$1.06\times 10^5$}&$4.05\times 10^5$&$1.00\times 10^6$&$2.88\times 10^3$&$6.94\times 10^4$&$5.56\times 10^3$&$\bm{2.69\times 10^3}$&$6.59\%$\\
				&Std.&\multicolumn{1}{|c}{$3.04\times 10^5$}&$4.05\times 10^5$&$7.97\times 10^2$&$4.15\times 10^2$&$2.53\times 10^5$&$7.90\times 10^2$&$\bm{3.28\times 10^2}$\\
				&Max&\multicolumn{1}{|c}{$1.00\times 10^6$}&$1.00\times 10^6$&$1.00\times 10^6$&$4.03\times 10^3$&$1.00\times 10^6$&$6.98\times 10^3$&$\bm{3.44\times 10^3}$\\
				&Min&\multicolumn{1}{|c}{$4.82\times 10^3$}&$4.63\times 10^3$&$1.00\times 10^6$&$2.41\times 10^3$&$2.29\times 10^3$&$3.96\times 10^3$&$\bm{1.94\times 10^3}$\\\bottomrule
		\end{tabular}}	
		\label{numerical results of USOP by MinUAV}
	\end{center}
	 \vspace{-0.6cm}
\end{table*}
\begin{table*}[t]
	\setlength{\abovedisplayskip}{1pt}
	\setlength{\belowdisplayskip}{1pt}
	\setlength{\abovecaptionskip}{1pt}
	\tiny
	\tabcolsep=0.15cm
	\begin{center}
		\setlength{\belowcaptionskip}{-0.2cm}
		\caption{Statistical results of MinAveEnergy obtained by NSGA-III-FDU and other multi-objective comparison approaches for NetResSOP}
		
		{\begin{tabular}{cccccccccc}\toprule
				&&\textbf{MOSPO} &\textbf{MOPSO} &\textbf {MOEA/D} &\textbf{NSGA-II} &\textbf{NSGA-III} &\textbf{IMSSA} &\textbf{NSGA-III-FDU}&\textbf{Improvement}\\\cmidrule(lr){3-10}&&\multicolumn{7}{c}{\textbf {Scale 1}}\\\cmidrule(lr){1-10}
				\multirow{4}{*}{$f_1$}
				&Mean&\multicolumn{1}{|c}{$7.09\times 10^5$}&$6.30\times 10^5$&$-1.95\times 10^6$&$1.43\times 10^6$&$1.48\times 10^6$&$1.49\times 10^6$&$\bm{1.93\times 10^6}$&$29.53\%$\\
				&Std.&\multicolumn{1}{|c}{$3.23\times 10^5$}&$2.99\times 10^5$&$4.33\times 10^6$&$5.00\times 10^5$&$4.56\times 10^5$&$\bm{2.65\times 10^5}$&$2.73\times 10^5$\\
				&Max&\multicolumn{1}{|c}{$1.34\times 10^6$}&$1.25\times 10^6$&$1.42\times 10^6$&$2.24\times 10^6$&$\bm{2.56\times 10^6}$&$2.01\times 10^6$&$2.22\times 10^6$\\
				&Min&\multicolumn{1}{|c}{$2.69\times 10^5$}&$1.35\times 10^5$&$-9.41\times 10^6$&$3.87\times 10^5$&$4.36\times 10^5$&$8.30\times 10^5$&$\bm{6.12\times 10^5} $\\\cmidrule(lr){1-10}
				\multirow{4}{*}{$f_2$}
				&Mean&\multicolumn{1}{|c}{5.13}&4.76&6.86&4.10&4.06&\textbf{4.03}&\textbf{4.03}&0.00\%\\
				&Std.&\multicolumn{1}{|c}{1.30}&0.89&4.15&0.40&0.25&\textbf{0.18}&\textbf{0.18}\\
				&Max&\multicolumn{1}{|c}{8.00}&7.00&15.00&6.00&\textbf{5.00}&\textbf{5.00}&\textbf{5.00}\\
				&Min&\multicolumn{1}{|c}{\textbf{4.00}}&\textbf{4.00}&\textbf{4.00}&\textbf{4.00}&\textbf{4.00}&\textbf{4.00}&\textbf{4.00}\\\cmidrule(lr){1-10}
				\multirow{4}{*}{$f_3$}
				&Mean&\multicolumn{1}{|c}{$3.29\times 10^3$}&$5.01\times 10^3$&$2.72\times 10^5$&$2.15\times 10^3$&$2.02\times 10^3$&$3.65\times 10^3$&$\bm{1.96\times 10^3}$&$2.97\%$\\
				&Std.&\multicolumn{1}{|c}{$1.08\times 10^3$}&$1.85\times 10^3$&$4.49\times 10^5$&$3.74\times 10^2$&$5.15\times 10^2$&$5.21\times 10^2$&$\bm{2.86\times 10^2}$\\
				&Max&\multicolumn{1}{|c}{$5.83\times 10^3$}&$8.94\times 10^3$&$1.00\times 10^6$&$3.29\times 10^3$&$2.96\times 10^3$&$4.50\times 10^3$&$\bm{2.43\times 10^3}$\\
				&Min&\multicolumn{1}{|c}{$1.67\times 10^3$}&$2.18\times 10^3$&$4.20\times 10^3$&$1.51\times 10^3$&$\bm{1.08\times 10^3}$&$2.41\times 10^3$&$1.35\times 10^3$\\\cmidrule(lr){1-10}
				&&\multicolumn{7}{c}{\textbf {Scale 2}}\\\cmidrule(lr){1-10}
				\multirow{4}{*}{$f_1$}
				&Mean&\multicolumn{1}{|c}{$1.18\times 10^6$}&$-2.00\times 10^6$&$-7.43\times 10^6$&$3.11\times 10^6$&$2.57\times 10^6$&$2.77\times 10^6$&$\bm{3.75\times 10^6}$&$20.57\%$\\
				&Std.&\multicolumn{1}{|c}{$3.13\times 10^6$}&$5.11\times 10^6$&$4.20\times 10^5$&$\bm{1.00\times 10^6}$&$2.44\times 10^6$&$6.85\times 10^5$&$1.22\times 10^6$\\
				&Max&\multicolumn{1}{|c}{$3.44\times 10^6$}&$2.83\times 10^6$&$-6.66\times 10^6$&$4.84\times 10^6$&$4.58\times 10^6$&$4.29\times 10^6$&$\bm{5.25\times 10^6}$\\
				&Min&\multicolumn{1}{|c}{$-8.34\times 10^5$}&$-8.75\times 10^6$&$-8.10\times 10^6$&$1.12\times 10^6$&$-6.26\times 10^6$&$1.60\times 10^6$&$\bm{1.75\times 10^6}$\\\cmidrule(lr){1-10}
				\multirow{4}{*}{$f_2$}
				&Mean&\multicolumn{1}{|c}{10.73}&12.00&20.20&8.46&8.83&8.30&\textbf{8.13}&2.04\%\\
				&Std.&\multicolumn{1}{|c}{3.59}&4.14&2.49&1.22&2.15&0.59&\textbf{0.34}\\
				&Max&\multicolumn{1}{|c}{22.00}&19.00&24.00&14.00&17.00&10.00&\textbf{9.00}\\
				&Min&\multicolumn{1}{|c}{\textbf{8.00}}&\textbf{8.00}&16.00&\textbf{8.00}&\textbf{8.00}&\textbf{8.00}&\textbf{8.00}\\\cmidrule(lr){1-10}
				\multirow{4}{*}{$f_3$}
				&Mean&\multicolumn{1}{|c}{$1.06\times 10^5$}&$4.04\times 10^5$&$1.00\times 10^6$&$2.75\times 10^3$&$6.93\times 10^4$&$5.03\times 10^3$&$\bm{2.61\times 10^3}$&$5.09\%$\\
				&Std.&\multicolumn{1}{|c}{$3.04\times 10^5$}&$4.97\times 10^5$&$6.36\times 10^2$&$4.06\times 10^2$&$2.53\times 10^5$&$4.43\times 10^2$&$\bm{3.40\times 10^2}$\\
				&Max&\multicolumn{1}{|c}{$1.00\times 10^6$}&$1.00\times 10^6$&$1.00\times 10^6$&$3.91\times 10^3$&{$1.00\times 10^6$}&$5.56\times 10^3$&$\bm{3.41\times 10^3}$\\
				&Min&\multicolumn{1}{|c}{$4.37\times 10^3$}&$4.43\times 10^3$&$1.00\times 10^6$&$2.28\times 10^3$&$2.20\times 10^3$&$3.73\times 10^3$&$\bm{1.86\times 10^3}$\\\bottomrule
		\end{tabular}}
		\label{numerical results of USOP by MinAveEnergy}
	\end{center}
	 \vspace{-0.4cm}
\end{table*}

\par Specifically, \emph{\textbf{(1) for Strategy MaxNetCap}}, under Scale 1, we can increase the D2D network capacity by $24.40\%$ approximately, while reducing the average energy consumption over all UAVs by $5.26\%$ approximately, when the number of deployed UAVs differs by less than $5\%$ on average. Besides, under Scale 2, the D2D network capacity is increased by $40.50\%$ approximately, and the average energy consumption is reduced by $5.51\%$ approximately, when using almost the same number of UAVs. \emph{\textbf{(2) For Strategy MinUAV}}, the gap on the number of deployed UAVs under Scale 1 is $0.00\%$, and Scale 2 shows similar results. However, other two optimization objectives can be optimized. Under Scale 1, the D2D network capacity is increased by $31.37\%$ approximately, while reducing the average energy consumption over all UAVs by $5.02\%$ approximately. Moreover, under Scale 2, we increase the network capacity by $37.46\%$ approximately, while the energy consumption will be saved by $6.59\%$. \emph{\textbf{(3) For Strategy MinAveEnergy}}, the gaps on the number of deployed UAVs are $0.00\%$ and $2.04\%$. However, the gaps on the D2D network capacity is $29.53\%$ and $20.57\%$ for Scales 1 and 2, respectively, while the energy consumptions are saved by $2.97\%$ and $5.09\%$. Note that all of abovementioned improvements are based on comparison of the optimization results obtained by NSGA-III-FDU and the suboptimal results obtained by other comparison algorithms, which means that the improvements could be even greater in actual deployments, since different comparison algorithms may obtain suboptimal solutions on different optimization objectives.

\section{Discussion}
\label{Discussion}
\subsection{Energy efficiency analysis with and without UAVs} 
\par In this part, we supply simulations to compare the network energy efficiency gaps with and without UAVs. As mentioned in Section \ref{Energy consumption model of UAV}, using UAVs as relays will generate extra flight energy consumption as well as communication energy consumption, and the energy consumption during flight will be far greater than the energy consumption of network communications of WDs and UAVs. Thus, it is unfair to compare the energy consumption during UAV flight with that of the network communications. Thus, to reflect the effectiveness of UAV relays, we compare the communication energy efficiency of the D2D network with and without UAVs. Specifically, for the case of using UAVs, we consider the total UAV transmission powers and SWD transmission powers. For the case without UAVs, we consider the direct transmission from SWD to DWD. Thus, the total SWD transmission powers are considered. To calculate the efficiency, we divide D2D network capacity, i.e., the first optimization objective $f_1$, by the abovementioned sum, and the results are listed in Table \ref{table:With_without_UAVs}.

\begin{table}[tbp]
	\setlength{\abovedisplayskip}{1pt}
	\setlength{\belowdisplayskip}{1pt}
	\setlength{\abovecaptionskip}{1pt}
	\tiny 
	\tabcolsep=0.3mm
	\vspace{-0.2cm}
	\begin{center}
		\caption{Energy efficiency and communication performance with and without UAVs obtained by NSGA-III-FDU}
		\vspace{-0.2cm}
		\begin{tabular}{ccccc}
			\toprule
			 & {$\begin{array}{l}\textbf { Communication } \\
			 		\textbf {energy efficiency}\\ \textbf { with UAVs [bit/J]}\end{array}$} &{$\begin{array}{l}\textbf { Communication } \\
			 		\textbf {energy efficiency}\\ \textbf { without UAVs [bit/J]}\end{array}$} &{$\begin{array}{l}\textbf { D2D network capacity} \\
			 		 \textbf { with UAVs [bps]}\end{array}$} &{$\begin{array}{l}\textbf { D2D network capacity} \\
			 		 \textbf { without UAVs [bps]}\end{array}$} 
		 		\\\cmidrule(l){2-5}  
			\textbf{Scale 1} & $\bm{9.68 \times 10^5}$ &$3.43 \times 10^5$ &$\bm{2.01 \times 10^6}$ &$3.43\times 10^4$\\
			\textbf{Scale 2} & $8.47 \times 10^5$ &$\bm{1.09 \times 10^6}$ &$\bm{5.07 \times 10^6}$ &$1.09 \times 10^6$\\ \bottomrule

		\end{tabular}
		\label{table:With_without_UAVs}
	\end{center}
	\vspace{-0.2cm}
\end{table}
\par As can be seen, the communication energy efficiency with UAVs in Scale 1 is better than that without UAVs, whereas the opposite result is obtained in Scale 2. The reason may be related to the differences in the transmission powers. Specifically, the transmission powers of SWDs are $0.01$ W, while the transmission powers of UAVs are $0.1 \sim 1$ W. In other words, the transmission powers of the UAVs may be dozens of times the power of the WDs. Moreover, the number of deployed UAVs in Scale 1 is less than that in Scale 2, which causes that the UAVs consume more transmission energy in Scale 2, such that obtaining a low communication energy efficiency. Although using UAVs may lose some communication energy efficiency, it improves D2D network capacities significantly in both Scales 1 and 2. Meanwhile, when flight energy consumption of UAVs is considered, the communication energy consumption is negligible. Thus, we can say that our proposed UAV-aided D2D communication is reasonable.
\subsection{Implementation analysis}
\par
\begin{figure}[tbp]
	\centering{\includegraphics[width=2.5in]{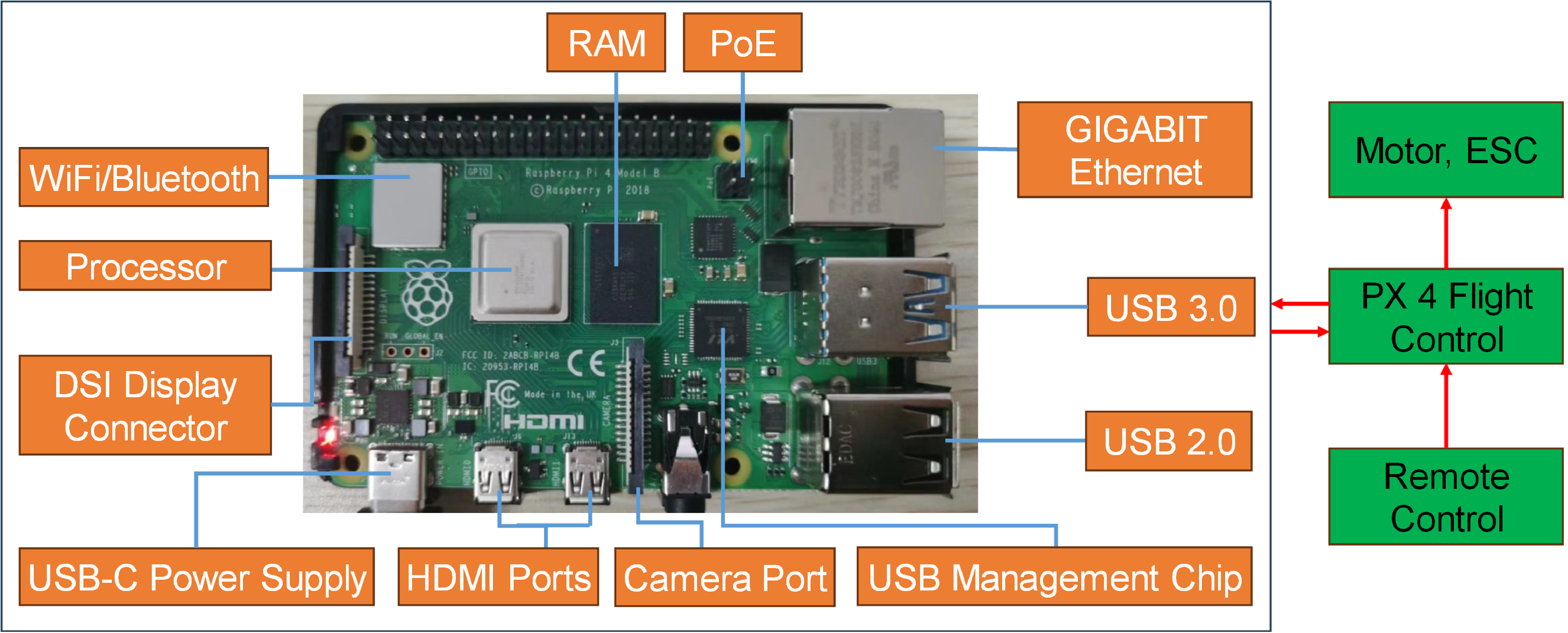}}
	\caption{Schematic diagram of autonomous UAV system based on Raspberry Pi.}
	\label{Implementation}
\end{figure}

 As mentioned above, the complexity of NSGA-III-FDU is acceptable. For further analyzing the implementability of the algorithm, we port NSGA-III-FDU to the Python platform in the Raspberry Pi 4B, which can be often regarded as the processor of many practical flight control systems of UAVs \cite{10012331} \cite{sun2023uav}. The schematic diagram of autonomous UAV system based on Raspberry Pi is shown in Fig. \ref{Implementation}, and several previous works have used it as the computational unit platform of UAVs \cite{10012331}. Moreover, several algorithms with good performance, which are NSGA-II, NSGA-III and IMSSA, are also ported to Python, to verify the implementability. Note that the objective functions are usually replaced by the lightweight proxy models in the actual deployment \cite{sun2023uav} \cite{jeong2005efficient}, since the proxy models can simulate the objective functions of complex problems, while returning numerical results quickly without extensive computation. Then, we set the parameters of Scale 1 to implement the algorithms. 
\par It can be seen from Table \ref{table:Implementation} that the execution time of NSGA-III-FDU is about $89.95$ s, which is a reasonable interval. The reasons can be summarized as follows: \textbf{a)} this work considers to simultaneously deploy all UAVs to serve the D2D network. Once one of the UAVs runs out of energy, all UAVs will be recalled, and then a new batch of UAVs will be deployed to achieve the service of seamless coverage. Thus, if the maximum hovering time of the UAV is greater than the computation time of the algorithm, the solution is reasonable, since the algorithm can be run in advance when the last batch of UAVs starts service. According to \cite{DBLP:journals/tits/MuntahaHJH21}, the UAV can usually hover for more than $15$ minutes in a dense urban, which is far beyond the required computation time.  \textbf{b)} The computation time when actually deployed will be less than the obtained results, since Python has a low code execution efficiency compared to C/C++. Thus, we can deploy C/C++ version of the algorithm to further improve the execution efficiency when actually deployed. Moreover, the computation power is also insignificant compared to the propulsion power of the UAVs \cite{10012331}, which can be easily tackled by the on-board energy. Accordingly, the proposed approach has good implementability in practical scenarios.

\begin{table}
	\setlength{\abovedisplayskip}{1pt}
	\setlength{\belowdisplayskip}{1pt}
	\setlength{\abovecaptionskip}{5pt}
	\tiny 
	\tabcolsep=1.5mm
	\begin{center}
		\caption{Execution time obtained by the proposed NSGA-III-FDU and several algorithms with good performance.}
		\begin{tabular}{ccccc}
			\toprule
			\textbf{Approach} & \textbf{NSGA-II} &\textbf{NSGA-III}&\textbf{IMSSA}&\textbf{NSGA-III-FDU}
			\\\cmidrule(l){2-5}  
			\textbf{Execution time [s]} & $\bm{15.33}$ &$54.82$ &$26.71$ &$89.85$\\ \bottomrule
			
		\end{tabular}
		\label{table:Implementation}
	\end{center}
\end{table}

\vspace{-0.2cm}
\section{Conclusion}
\label{Conclusion}
\par In this paper, the resource scheduling in UAVs-aided D2D networks is investigated. First, we employ multiple UAVs as aerial relays to serve the UAV-aided D2D pairs subject to the direct D2D pairs. Then, an NetResSOP is formulated to jointly consider the number of deployed UAVs, UAV positions, UAV transmission powers, UAV flight velocities, communication channels, and UAV-device pair assignment so as to maximize the D2D network capacity, minimize the number of deployed UAVs, and minimize the average energy consumption over all UAVs simultaneously. Since NetResSOP is an MIPP as well as an NP-hard problem, we propose an NSGA-III-FDU to solve the problem comprehensively in polynomial time. Simulations results demonstrate that NSGA-III-FDU outperforms other comparison strategies and approaches, such as UD, RD, MGO, MOSOP, MOPSO, MOEA/D, NSGA-II, NSGA-III and IMSSA. Moreover, the effectiveness of the proposed approach is verified, and it is more stable under all alternative strategies.
\vspace{-0.6cm}


%

\ifCLASSOPTIONcaptionsoff
  \newpage
\fi



%
%

\bibliographystyle{ieeetr}
\bibliography{ref-MyUAV}

\begin{thebibliography}{10}

\bibitem{DBLP:journals/corr/abs-2202-06046}
Z.~Jia, Q.~Wu, C.~Dong, C.~Yuen, and Z.~Han, ``Hierarchical aerial computing
  for {Internet of Things} via cooperation of {HAPs} and {UAVs},'' {\em {IEEE}
  Internet Things J.}, vol.~10, no.~7, pp.~5676--5688, 2023.

\bibitem{DBLP:journals/tvt/GaoJXYFL22}
H.~Gao, C.~Jia, W.~Xu, C.~Yuen, Z.~Feng, and Y.~Lu, ``Machine
  learning-empowered beam management for {mmWave-NOMA} in multi-{UAVs}
  networks,'' {\em {IEEE} Trans. Veh. Technol.}, vol.~71, no.~8,
  pp.~8487--8502, 2022.

\bibitem{DBLP:journals/iotj/LiuP0WLL22}
Y.~Liu, H.~Pan, G.~Sun, A.~Wang, J.~Li, and S.~Liang, ``Joint scheduling and
  trajectory optimization of charging {UAV} in wireless rechargeable sensor
  networks,'' {\em {IEEE} Internet Things J.}, vol.~9, no.~14,
  pp.~11796--11813, 2022.

\bibitem{DBLP:journals/iotj/XuLHY21}
Y.~Xu, Z.~Liu, C.~Huang, and C.~Yuen, ``Robust resource allocation algorithm
  for energy-harvesting-based {D2D} communication underlaying {UAV}-assisted
  networks,'' {\em {IEEE} Internet Things J.}, vol.~8, no.~23,
  pp.~17161--17171, 2021.

\bibitem{DBLP:journals/ton/ZhongGLC20}
X.~Zhong, Y.~Guo, N.~Li, and Y.~Chen, ``Joint optimization of relay deployment,
  channel allocation, and relay assignment for {UAVs}-aided {D2D} networks,''
  {\em {IEEE/ACM} Trans. Netw.}, vol.~28, no.~2, pp.~804--817, 2020.

\bibitem{DBLP:conf/vtc/JavidsharifiAKS22}
M.~Javidsharifi, H.~P. Arabani, T.~Kerekes, D.~Sera, and J.~M. Guerrero,
  ``{PV}-powered base stations equipped by {UAVs} in urban areas,'' in {\em
  96th Vehicular Technology Conference, {VTC} Fall 2022, London, United
  Kingdom, September 26-29, 2022}, pp.~1--4, {IEEE}, 2022.

\bibitem{DBLP:journals/tcom/HanBBC22}
C.~Han, L.~Bai, T.~Bai, and J.~Choi, ``Joint {UAV} deployment and power
  allocation for secure space-air-ground communications,'' {\em {IEEE} Trans.
  Commun.}, vol.~70, no.~10, pp.~6804--6818, 2022.

\bibitem{DBLP:journals/icl/KaleemKMNYK22}
Z.~Kaleem, W.~Khalid, A.~H. Muqaibel, A.~A. Nasir, C.~Yuen, and G.~K.
  Karagiannidis, ``Learning-aided {UAV} {3D} placement and power allocation for
  sum-capacity enhancement under varying altitudes,'' {\em {IEEE} Commun.
  Lett.}, vol.~26, no.~7, pp.~1633--1637, 2022.

\bibitem{DBLP:journals/tcom/SunLWWSL22}
G.~Sun, J.~Li, A.~Wang, Q.~Wu, Z.~Sun, and Y.~Liu, ``Secure and
  energy-efficient {UAV} relay communications exploiting collaborative
  beamforming,'' {\em {IEEE} Trans. Commun.}, vol.~70, no.~8, pp.~5401--5416,
  2022.

\bibitem{DBLP:journals/tifs/ChenLWHXZ23}
H.~Chen, H.~Li, Y.~Wang, M.~Hao, G.~Xu, and T.~Zhang, ``{PriVDT}: {An}
  efficient two-party cryptographic framework for vertical decision trees,''
  {\em {IEEE} Trans. Inf. Forensics Secur.}, vol.~18, pp.~1006--1021, 2023.

\bibitem{DBLP:journals/iotj/JiZNW21}
J.~Ji, K.~Zhu, D.~Niyato, and R.~Wang, ``Joint trajectory design and resource
  allocation for secure transmission in cache-enabled {UAV}-relaying networks
  with {D2D} communications,'' {\em {IEEE} Internet Things J.}, vol.~8, no.~3,
  pp.~1557--1571, 2021.

\bibitem{DBLP:journals/tcom/AlsharoaY21}
A.~Alsharoa and M.~Yuksel, ``Energy efficient {D2D} communications using
  multiple {UAV} relays,'' {\em {IEEE} Trans. Commun.}, vol.~69, no.~8,
  pp.~5337--5351, 2021.

\bibitem{DBLP:journals/tcom/SuPCJZY22}
Y.~Su, X.~Pang, S.~Chen, X.~Jiang, N.~Zhao, and F.~R. Yu, ``Spectrum and energy
  efficiency optimization in {IRS}-assisted {UAV} networks,'' {\em {IEEE}
  Trans. Commun.}, vol.~70, no.~10, pp.~6489--6502, 2022.

\bibitem{DBLP:journals/iotj/GulEK22}
O.~M. Gul, A.~M. Erkmen, and B.~Kantarci, ``{UAV}-driven sustainable and
  quality-aware data collection in robotic wireless sensor networks,'' {\em
  {IEEE} Internet Things J.}, vol.~9, no.~24, pp.~25150--25164, 2022.

\bibitem{DBLP:journals/tits/HanZZL22}
S.~Han, K.~Zhu, M.~Zhou, and X.~Liu, ``Joint deployment optimization and flight
  trajectory planning for {UAV} assisted iot data collection: {A} bilevel
  optimization approach,'' {\em {IEEE} Trans. Intell. Transp. Syst.}, vol.~23,
  no.~11, pp.~21492--21504, 2022.

\bibitem{DBLP:journals/twc/WangDZ19}
Z.~Wang, L.~Duan, and R.~Zhang, ``Adaptive deployment for {UAV}-aided
  communication networks,'' {\em {IEEE} Trans. Wirel. Commun.}, vol.~18, no.~9,
  pp.~4531--4543, 2019.

\bibitem{DBLP:journals/tmc/ZhangD19}
X.~Zhang and L.~Duan, ``Fast deployment of {UAV} networks for optimal wireless
  coverage,'' {\em {IEEE} Trans. Mob. Comput.}, vol.~18, no.~3, pp.~588--601,
  2019.

\bibitem{DBLP:journals/tvt/ZhongGLL20}
X.~Zhong, Y.~Guo, N.~Li, and S.~Li, ``Joint relay assignment and channel
  allocation for opportunistic {UAVs}-aided dynamic networks: {A} mood-driven
  approach,'' {\em {IEEE} Trans. Veh. Technol.}, vol.~69, no.~12,
  pp.~15019--15034, 2020.

\bibitem{DBLP:journals/tcom/ZhangZZZXX21}
C.~Zhang, L.~Zhang, L.~Zhu, T.~Zhang, Z.~Xiao, and X.~Xia, ``3{D} deployment of
  multiple {UAV}-mounted base stations for {UAV} communications,'' {\em {IEEE}
  Trans. Commun.}, vol.~69, no.~4, pp.~2473--2488, 2021.

\bibitem{DBLP:journals/tcom/LiuHZZLL22}
Y.~Liu, W.~Huangfu, H.~Zhou, H.~Zhang, J.~Liu, and K.~Long, ``Fair and
  energy-efficient coverage optimization for {UAV} placement problem in the
  cellular network,'' {\em {IEEE} Trans. Commun.}, vol.~70, no.~6,
  pp.~4222--4235, 2022.

\bibitem{DBLP:journals/tgcn/ZhangOCWMC22}
G.~Zhang, X.~Ou, M.~Cui, Q.~Wu, S.~Ma, and W.~Chen, ``Cooperative {UAV} enabled
  relaying systems: Joint trajectory and transmit power optimization,'' {\em
  {IEEE} Trans. Green Commun. Netw.}, vol.~6, no.~1, pp.~543--557, 2022.

\bibitem{DBLP:journals/iotj/LiZZY21}
B.~Li, S.~Zhao, R.~Zhang, and L.~Yang, ``Full-duplex {UAV} relaying for
  multiple user pairs,'' {\em {IEEE} Internet Things J.}, vol.~8, no.~6,
  pp.~4657--4667, 2021.

\bibitem{DBLP:journals/tgcn/LiuCZWCZ21}
T.~Liu, M.~Cui, G.~Zhang, Q.~Wu, X.~Chu, and J.~Zhang, ``3{D} trajectory and
  transmit power optimization for {UAV}-enabled multi-link relaying systems,''
  {\em {IEEE} Trans. Green Commun. Netw.}, vol.~5, no.~1, pp.~392--405, 2021.

\bibitem{DBLP:journals/iotj/FengWCWGL19}
W.~Feng, J.~Wang, Y.~Chen, X.~Wang, N.~Ge, and J.~Lu, ``{UAV}-aided {MIMO}
  communications for 5{G} internet of things,'' {\em {IEEE} Internet Things
  J.}, vol.~6, no.~2, pp.~1731--1740, 2019.

\bibitem{DBLP:journals/wcl/Al-HouraniSL14}
A.~Al{-}Hourani, K.~Sithamparanathan, and S.~Lardner, ``Optimal {LAP} altitude
  for maximum coverage,'' {\em {IEEE} Wirel. Commun. Lett.}, vol.~3, no.~6,
  pp.~569--572, 2014.

\bibitem{DBLP:journals/twc/ZengXZ18}
Y.~Zeng, X.~Xu, and R.~Zhang, ``Trajectory design for completion time
  minimization in {UAV}-enabled multicasting,'' {\em {IEEE} Trans. Wirel.
  Commun.}, vol.~17, no.~4, pp.~2233--2246, 2018.

\bibitem{DBLP:journals/twc/ZengXZ19}
Y.~Zeng, J.~Xu, and R.~Zhang, ``Energy minimization for wireless communication
  with rotary-wing {UAV},'' {\em {IEEE} Trans. Wirel. Commun.}, vol.~18, no.~4,
  pp.~2329--2345, 2019.

\bibitem{DBLP:journals/pieee/ZengWZ19}
Y.~Zeng, Q.~Wu, and R.~Zhang, ``Accessing from the sky: {A} tutorial on {UAV}
  communications for 5{G} and beyond,'' {\em Proc. {IEEE}}, vol.~107, no.~12,
  pp.~2327--2375, 2019.

\bibitem{DBLP:journals/tit/LanemanTW04}
J.~N. Laneman, D.~N.~C. Tse, and G.~W. Wornell, ``Cooperative diversity in
  wireless networks: Efficient protocols and outage behavior,'' {\em {IEEE}
  Trans. Inf. Theory}, vol.~50, no.~12, pp.~3062--3080, 2004.

\bibitem{DBLP:journals/candie/MengZRZL20}
L.~Meng, C.~Zhang, Y.~Ren, B.~Zhang, and C.~Lv, ``Mixed-integer linear
  programming and constraint programming formulations for solving distributed
  flexible job shop scheduling problem,'' {\em Comput. Ind. Eng.}, vol.~142,
  p.~106347, 2020.

\bibitem{nair2020solving}
V.~Nair, S.~Bartunov, F.~Gimeno, I.~Von~Glehn, P.~Lichocki, I.~Lobov,
  B.~O'Donoghue, N.~Sonnerat, C.~Tjandraatmadja, P.~Wang, {\em et~al.},
  ``Solving mixed integer programs using neural networks,'' {\em arXiv preprint
  arXiv:2012.13349}, 2020.

\bibitem{DBLP:journals/tmc/WangYHGX18}
L.~Wang, Z.~Yu, Q.~Han, B.~Guo, and H.~Xiong, ``Multi-objective optimization
  based allocation of heterogeneous spatial crowdsourcing tasks,'' {\em {IEEE}
  Trans. Mob. Comput.}, vol.~17, no.~7, pp.~1637--1650, 2018.

\bibitem{9777886}
H.~Pan, Y.~Liu, G.~Sun, J.~Fan, L.~Shuang, and C.~Yuen, ``Joint power and {3D}
  trajectory optimization for {UAV}-enabled wireless powered communication
  networks with obstacles,'' {\em {IEEE} Trans. Commun.}, pp.~1--1, 2023.

\bibitem{zhao2023self}
J.~Zhao, Z.~Wang, J.~Cao, and K.~H. Cheong, ``A self-adaptive evolutionary
  deception framework for community structure,'' {\em {IEEE} Trans. Syst. Man
  Cybern. Syst.}, 2023.

\bibitem{DBLP:journals/iotj/LiuWSL22}
L.~Liu, A.~Wang, G.~Sun, and J.~Li, ``Multiobjective optimization for improving
  throughput and energy efficiency in {UAV}-enabled {IoT},'' {\em {IEEE}
  Internet Things J.}, vol.~9, no.~20, pp.~20763--20777, 2022.

\bibitem{DBLP:journals/tec/DebJ14}
K.~Deb and H.~Jain, ``An evolutionary many-objective optimization algorithm
  using reference-point-based nondominated sorting approach, part {I:} solving
  problems with box constraints,'' {\em {IEEE} Trans. Evol. Comput.}, vol.~18,
  no.~4, pp.~577--601, 2014.

\bibitem{DBLP:journals/tec/JainD14}
H.~Jain and K.~Deb, ``An evolutionary many-objective optimization algorithm
  using reference-point based nondominated sorting approach, part {II:}
  handling constraints and extending to an adaptive approach,'' {\em {IEEE}
  Trans. Evol. Comput.}, vol.~18, no.~4, pp.~602--622, 2014.

\bibitem{DBLP:journals/tec/DebAPM02}
K.~Deb, S.~Agrawal, A.~Pratap, and T.~Meyarivan, ``A fast and elitist
  multiobjective genetic algorithm: {NSGA-II},'' {\em {IEEE} Trans. Evol.
  Comput.}, vol.~6, no.~2, pp.~182--197, 2002.

\bibitem{DBLP:conf/cec/Zhao08}
X.~Zhao, ``Convergent analysis on evolutionary algorithm with non-uniform
  mutation,'' in {\em Proceedings of the {IEEE} Congress on Evolutionary
  Computation, {CEC} 2008, June 1-6, 2008, Hong Kong, China}, pp.~940--944,
  {IEEE}, 2008.

\bibitem{DBLP:journals/isci/ChenH21}
Y.~Chen and J.~He, ``Average convergence rate of evolutionary algorithms in
  continuous optimization,'' {\em Inf. Sci.}, vol.~562, pp.~200--219, 2021.

\bibitem{DBLP:journals/tec/HeL16}
J.~He and G.~Lin, ``Average convergence rate of evolutionary algorithms,'' {\em
  {IEEE} Trans. Evol. Comput.}, vol.~20, no.~2, pp.~316--321, 2016.

\bibitem{DBLP:journals/tec/HuynhCSR18}
Q.~N. Huynh, S.~Chand, H.~K. Singh, and T.~Ray, ``Genetic programming with
  mixed-integer linear programming-based library search,'' {\em {IEEE} Trans.
  Evol. Comput.}, vol.~22, no.~5, pp.~733--747, 2018.

\bibitem{DBLP:journals/twc/LuWCCFL18}
J.~Lu, S.~Wan, X.~Chen, Z.~Chen, P.~Fan, and K.~B. Letaief, ``Beyond empirical
  models: Pattern formation driven placement of {UAV} base stations,'' {\em
  {IEEE} Trans. Wirel. Commun.}, vol.~17, no.~6, pp.~3641--3655, 2018.

\bibitem{10012331}
J.~Li, G.~Sun, H.~Kang, A.~Wang, S.~Liang, Y.~Liu, and Y.~Zhang,
  ``Multi-objective optimization approaches for physical layer secure
  communications based on collaborative beamforming in {UAV} networks,'' {\em
  {IEEE/ACM} Trans. Netw.}, pp.~1--16, 2023.

\bibitem{abdollahzadeh2022mountain}
B.~Abdollahzadeh, F.~S. Gharehchopogh, N.~Khodadadi, and S.~Mirjalili,
  ``Mountain gazelle optimizer: {A} new nature-inspired metaheuristic algorithm
  for global optimization problems,'' {\em Adv. Eng. Softw.}, vol.~174,
  p.~103282, 2022.

\bibitem{khodadadi2022multi}
N.~Khodadadi, L.~Abualigah, and S.~Mirjalili, ``Multi-objective stochastic
  paint optimizer {(MOSPO)},'' {\em Neural. Comput. Appl.}, vol.~34, no.~20,
  pp.~18035--18058, 2022.

\bibitem{DBLP:journals/tec/ZhangL07}
Q.~Zhang and H.~Li, ``{MOEA/D:} {A} multiobjective evolutionary algorithm based
  on decomposition,'' {\em {IEEE} Trans. Evol. Comput.}, vol.~11, no.~6,
  pp.~712--731, 2007.

\bibitem{DBLP:journals/isci/TripathiBP07}
P.~K. Tripathi, S.~Bandyopadhyay, and S.~K. Pal, ``Multi-objective particle
  swarm optimization with time variant inertia and acceleration coefficients,''
  {\em Inf. Sci.}, vol.~177, no.~22, pp.~5033--5049, 2007.

\bibitem{DBLP:journals/tcyb/ShenWW22}
J.~Shen, P.~Wang, and X.~Wang, ``A controlled strengthened dominance relation
  for evolutionary many-objective optimization,'' {\em {IEEE} Trans. Cybern.},
  vol.~52, no.~5, pp.~3645--3657, 2022.

\bibitem{sun2023uav}
G.~Sun, X.~Zheng, Z.~Sun, Q.~Wu, J.~Li, Y.~Liu, and V.~C. Leung,
  ``{UAV}-enabled secure communications via collaborative beamforming with
  imperfect eavesdropper information,'' {\em {IEEE} Trans. Mob. Comput.}, 2023.

\bibitem{DBLP:journals/twc/MehariPCDVPJMDM16}
M.~T. Mehari, E.~D. Poorter, I.~Couckuyt, D.~Deschrijver, G.~Vermeeren,
  D.~Plets, W.~Joseph, L.~Martens, T.~Dhaene, and I.~Moerman, ``Efficient
  identification of a multi-objective pareto front on a wireless
  experimentation facility,'' {\em {IEEE} Trans. Wirel. Commun.}, vol.~15,
  no.~10, pp.~6662--6675, 2016.

\bibitem{jeong2005efficient}
S.~Jeong, M.~Murayama, and K.~Yamamoto, ``Efficient optimization design method
  using kriging model,'' {\em J Aircr}, vol.~42, no.~2, pp.~413--420, 2005.

\bibitem{DBLP:journals/tits/MuntahaHJH21}
S.~T. Muntaha, S.~A. Hassan, H.~Jung, and M.~S. Hossain, ``Energy efficiency
  and hover time optimization in {UAV}-based {HetNets},'' {\em {IEEE} Trans.
  Intell. Transp. Syst.}, vol.~22, no.~8, pp.~5103--5111, 2021.

\end{thebibliography}

%




\end{document}